\documentclass[11pt,noindent]{article}
\usepackage{graphics,cite,amssymb,float,ifpdf}
\usepackage{amsmath, mathbbol}

\usepackage{amsfonts}

\usepackage{epsf}
\usepackage{epsfig}

\ifpdf
\usepackage[pdftex]{graphicx}
\usepackage[pdftex,dvipsnames]{xcolor}
\usepackage[final]{pdfpages}
\else
\usepackage{graphicx}
\usepackage[dvipsnames]{xcolor}
\usepackage{psfrag}
\fi
\ifpdf
\DeclareGraphicsExtensions{.pdf, .jpg, .png}
\else
\DeclareGraphicsExtensions{.eps, .jpg}
\fi

\usepackage{array}
\usepackage{multirow}
\def\lsim{\;\raise0.3ex\hbox{$<$\kern-0.75em\raise-1.1ex\hbox{$\sim$}}\;}
\def\gsim{\;\raise0.3ex\hbox{$>$\kern-0.75em\raise-1.1ex\hbox{$\sim$}}\;}

\def\ben{\begin{enumerate}}  \def\een{\end{enumerate}}
\def\bit{\begin{itemize}}    \def\eit{\end{itemize}}
\def\beq{\begin{equation}}   \def\eeq{\end{equation}}
\def\ba{\begin{array}}       \def\ea{\end{array}}
\def\bea{\begin{eqnarray}}   \def\eea{\end{eqnarray}}
\def\nn{\nonumber}

\renewcommand{\b}{{\mathbb B}}
\renewcommand{\c}{{\mathbb C}}
\newcommand{\cbar}{\bar{\mathbb C}}

\newcolumntype{C}{ >{\centering\arraybackslash} m{4cm} }

\begin{document}

\setcounter{footnote}{0}
\vspace*{-1.5cm}
\begin{flushright}
LPT Orsay 14-78 \\
PCCF RI 14-07

\vspace*{2mm}
\today								
\end{flushright}
\begin{center}
\vspace*{5mm}

\vspace{1cm}
{\Large\bf 
Indirect searches for sterile neutrinos at a high-luminosity 
\vspace*{0.1cm}
\\

$\boldsymbol{Z}$-factory}
\vspace*{0.8cm}

{\bf A. Abada$^{a}$, V. De Romeri$^{b}$, S. Monteil$^{b}$,
  J. Orloff$^{b}$ and A.~M. Teixeira$^{b}$} 
  
\vspace*{.5cm} 
$^{a}$ Laboratoire de Physique Th\'eorique, CNRS -- UMR 8627, \\
Universit\'e de Paris-Sud, F-91405 Orsay Cedex, France
\vspace*{.2cm} 

$^{b}$ Laboratoire de Physique Corpusculaire, CNRS/IN2P3 -- UMR 6533,\\ 
Campus des C\'ezeaux, 24 Av. des Landais, F-63177 Aubi\`ere Cedex, France
\end{center}

\vspace*{6mm}
\begin{abstract}
A future high-luminosity $Z$-factory will offer the possibility to
study rare $Z$ decays, as those leading to lepton flavour violating final states.
Processes such as $Z \to \ell_1^\mp \ell_2^\pm$
are potentially complementary to low-energy 
(high-intensity) observables of lepton flavour violation.
In this work we address the impact of new sterile fermions 
on lepton flavour violating $Z$ decays, focusing on potential searches
at FCC-ee (TLEP), and taking into account experimental and
observational constraints on the sterile states. We consider a minimal
extension of the Standard Model by one sterile fermion
state, and two well-motivated frameworks of neutrino mass generation,
the Inverse Seesaw embedded into the Standard Model, and the $\nu$MSM. 
Our study shows that sterile neutrinos can 
give rise to contributions to BR($Z \to \ell_1^\mp \ell_2^\pm$) 
within reach of the FCC-ee. We also discuss the complementarity between 
a high-luminosity $Z$-factory and low-energy charged lepton flavour
violation facilities.
\end{abstract}
\vspace*{4mm}

\section{Introduction}
\label{sec:intro}

Rare flavour-violating $Z$ decays, as is the case of those 
violating lepton flavour conservation 
$Z \to  e \mu,\, \mu\tau,\, e \tau $, provide a clear 
evidence for new physics beyond the Standard Model (SM).  
In the SM,  these decays are forbidden due to the GIM
mechanism~\cite{Glashow:1970st},  
and their rates remain extremely small (below $10^{-50}$) when the SM
is minimally (ad-hoc)  
extended to incorporate flavour violation in the neutral lepton sector
(neutrino masses and 
mixings)~\cite{Riemann:1982rq,Mann:1984dvt,Riemann:1999ab,Illana:1999ww}.  

Sizeable rates for  $Z\to \ell_1^{\mp}\ell_2^{\pm}$ processes 
reflect the existence of new particles, either coupling with sub-weak 
strength to the SM particles, or then sufficiently heavy to have
escaped direct detection at current high-energy searches. 
Among these feebly interacting particles, potentially at the origin of
$Z \to \ell_1^{\mp}\ell_2^{\pm}$  
decays, are sterile (gauge-neutral) fermions, arising in several minimal
extensions of the SM, as for instance in those aiming at 
addressing the origin of neutrino masses and mixings. The existence of
sterile states is further  
supported by current data from neutrino experiments 
(Gallium~\cite{gallium:I}, reactor~\cite{reactor:I} and
accelerator~\cite{Aguilar:2001ty,miniboone:I} anomalies).
Sterile neutrinos are also a popular solution for the dark matter (DM)
problem~\cite{Dodelson:1993je, Abazajian:2001nj,nu_WDM}, and
can potentially alleviate some tensions regarding structure formation
observations~\cite{general_structure}. (Although there is still a
tension between the most recent Planck results on extra light neutrinos
(relics) and reactor anomalies, in this work we focus on the r\^ole
of (heavier) sterile fermions, which are not expected to 
contribute as light relativistic degrees of freedom~\cite{Ade:2013zuv}.)  
Rare {\emph {charged}} lepton flavour violating (cLFV) $Z$ decays have been
extensively discussed in the context of SM extensions involving
massive (Majorana and/or Dirac)
neutrinos~\cite{Mann:1983dv,Illana:1999ww,Illana:2000ic,Ilakovac:1994kj};
similar studies were carried using an effective theory
approach~\cite{Perez:2003ad,FloresTlalpa:2001sp,Delepine:2001di,Davidson:2012wn}, 
some also exploring a possible complementarity with low-energy
cLFV searches. 

The current bounds on the branching ratios (BRs) for cLFV $Z$ decays, 
\begin{equation}\label{eq:BR_LFVZ}
{\rm BR}(Z\to \ell_1^{\mp}\ell_2^{\pm}) \,=\,
\frac{\Gamma(Z \to \ell_1^\pm \ell_2^\mp )} {\Gamma_Z}\,,
\end{equation}
were established by LEP, performing as a $Z$
factory; recently, the  
ATLAS experiment established new bounds on the corresponding BRs, 
significantly improving  the bound for $e\mu$ final states:
\begin{align}
&
{\rm BR}(Z\to e^{\mp}\mu^{\pm}) \,<\, 7.5 \times 10^{-7}~~ 
\mbox{\cite{Aad:2014bca}}\,, \\
&
{\rm BR}(Z\to e^{\mp}\tau^{\pm}) \,<\, 9.8 \times 10^{-6} ~~
\mbox{\cite{Adriani:1993sy,Akers:1995gz}}\,, \quad
{\rm BR}(Z\to \mu^{\mp}\tau^{\pm}) \,<\, 1.2 \times 10^{-5} ~~
\mbox{\cite{Akers:1995gz,Abreu:1996mj}}\,.
\end{align}

A future circular collider, running in electron-positron mode, FCC-ee
(TLEP)~\cite{FCC-WG}, will cons\-titute a true  
high-luminosity $Z$ factory, with an expected production of $10^{12}$
$Z$ bosons ($10^{13}$ with the ``crab-waist''), when operating at the $Z$ mass pole. 
Such large statistics (above Tera-$Z$) will thus allow to better determine the 
properties of the $Z$ boson, and to probe new physics (NP) scenarios through 
the above cLFV processes. 
The clean  nature of the cLFV $Z \to \ell_1^\mp \ell_2^\pm$ decays
(only charged leptons - especially muons - in the final state)
implies that the sensitivity to these rare processes is essentially
only constrained by the  
expected luminosity; one can thus foresee a significant improvement
in the experimental sensitivity at FCC-ee to rare cLFV $Z$ decays,  
for instance, $\text{BR}(Z\to e^{\mp}\mu^{\pm}) \,\sim \, 10^{-13}$.

Revisiting cLFV $Z$ decays in the presence of 
extra sterile fermions is particularly
timely given the present experimental context: not only we have 
reached an unprecedented precision in the determination of several
neutrino oscillation
parameters~\cite{Tortola:2012te,Fogli:2012ua,GonzalezGarcia:2012sz,Forero:2014bxa,nufit,Gonzalez-Garcia:2014bfa}, 
and new bounds on low-energy cLFV observables (for instance
MEG~\cite{Adam:2013mnn}), 
but we are also entering a challenging era, where many ambitious
(post-LHC) experimental projects   
are being put forward. 
Given their r\^ole in a vast array of observables (see, for 
instance~\cite{Abada:2012mc,Abada:2013aba,Abada:2014nwa} and
references therein), sterile neutrinos  
are becoming strong candidates for the physics case of several
post-LHC facilities, as is the case of the FCC-ee (TLEP). 
It is also worth mentioning that direct searches for (nearly) sterile fermions, 
as right-handed (RH) neutrinos, 
relying on their comparatively long lifetime, have recently been
studied in the context of  
high-luminosity $Z$-factories like the FCC-ee~\cite{Blondel:2014bra}.

The present work focuses on the potential of the FCC-ee to explore the
r\^ole  of cLFV  decays  of the $Z$ boson 
as indirect probes of sterile fermions~\cite{FCC-WG}, emphasising the
complementarity of these searches with  
respect to low-energy cLFV observables such as $\mu \to e \gamma$ and 
$\mu\to e e e$ decays and $\mu-e$ conversion in nuclei. 
We consider SM extensions via sterile neutrinos, with 
a non-negligible mixing to the light (mostly) active neutrinos, 
for a wide range of masses of the sterile mass spectrum. In
particular, we address three scenarios: a simple toy-model
extension of the SM with one sterile fermion (the ``3+1 model"), and two
well motivated frameworks for neutrino mass generation, the
$\nu$MSM~\cite{Asaka:2005an} and one realisation of the Inverse
Seesaw~\cite{Mohapatra:1986bd}. 

Our analysis (conducted for each of the above mentioned 
scenarios, which are confronted to all observational and 
experimental constraints, especially those from low-energy 
cLFV observables), reveals that sterile neutrinos can indeed give rise to 
contributions to BR($Z\to \ell_1^\mp \ell_2^\pm$)
within reach of the FCC-ee.

This work is organised as follows: in Section~\ref{sec:Zdecays:gen} we
consider the  
general formulation of the lepton flavour violating (and lepton
flavour conserving) 
BR($Z \to \ell^\mp \ell^\pm$) in terms of the sterile masses and mixings to
the active neutrinos.  
We also discuss the experimental prospects. In
Section~\ref{sec:constraints} we motivate this  
class of extensions and discuss the different observational (mainly
the cosmological ones)  and experimental bounds on  
sterile states. In the following three sections, we describe and
discuss in detail the prospects  
of different extensions of the SM regarding $Z \to \ell_1^\mp \ell_2^\pm$
decays at a high-luminosity $Z$  
factory, also addressing the complementarity with respect to other
low-energy observables.

\section{Leptonic $\boldsymbol{Z}$ decays in the presence of sterile
  neutrinos}\label{sec:Zdecays:gen} 
In the original formulation of the SM with massless
neutrinos and no mixing in the lepton sector, the couplings of the
gauge bosons to neutral and charged leptons are strictly flavour
conserving, lepton-flavour changing $Z$ decays being forbidden
due to the GIM mechanism~\cite{Glashow:1970st}.
Moreover, the couplings are flavour universal, so that in
the SM one has 
$g_{\ell_i \nu_i W} \propto g_w$, 
$g_{\ell_i \ell_i Z} \propto g_w$, as well as
$g_{\nu_i \nu_i Z} \propto g_w$, 
where $g_w$ denotes the weak coupling constant. 
These rates 
remain extremely small even in the case in which the SM is ``ad-hoc''-extended to incorporate three massive and  mixing
neutrinos~\cite{Riemann:1982rq,Mann:1984dvt,Ilakovac:1994kj,Riemann:1999ab,Illana:1999ww}: 
\begin{align}\label{old-tm}
\text{BR}(Z\to \mu^{\mp}\tau^{\pm}) \sim 10^{-54}\,, \quad 
\text{BR}(Z\to e^{\mp} \mu^{\pm}) \sim
\text{BR}(Z\to e^{\mp}\tau^{\pm}) \lsim  4 \times 10^{-60}\,.
\end{align}
Let us now consider the extension of the SM via $n_S$ additional 
sterile neutral (Majorana) fermions, 
mixing with
the active neutrinos.
In the physical lepton (or mass) basis, the SM Lagrangian is modified as follows%
\footnote{
See e.g.~\cite{Ilakovac:1994kj} for a detailed derivation starting from
explicit lepton mass matrices.
}: 
\begin{align}\label{eq:lagrangian:WGHZ}
& \mathcal{L}_{W^\pm}\, =\, -\frac{g_w}{\sqrt{2}} \, W^-_\mu \,
\sum_{l=1}^{3} \sum_{j=1}^{3 + n_S} {\bf U}_{lj} \bar \ell_l 
\gamma^\mu P_L \nu_j \, + \, \text{h.c.}\,, \nonumber \\
& \mathcal{L}_{Z^0}\, = \,-\frac{g_w}{2 \cos \theta_w} \, Z_\mu \,
\sum_{i,j=1}^{3 + n_S} \bar \nu_i \gamma ^\mu \left(
P_L {\bf C}_{ij} - P_R {\bf C}_{ij}^* \right) \nu_j\,, \nonumber \\
& \mathcal{L}_{H^0}\, = \, -\frac{g_w}{2 M_W} \, H  \,
\sum_{i,j=1}^{3 + n_S}  {\bf C}_{ij}  \bar \nu_i\left(
P_R m_i + P_L m_j \right) \nu_j + \, \text{h.c.}\, \nonumber \\
& \mathcal{L}_{G^0}\, =\,\frac{i g_w}{2 M_W} \, G^0 \,
\sum_{i,j=1}^{3 + n_S} {\bf C}_{ij}  \bar \nu_i  
\left(P_R m_j  - P_L m_i  \right) \nu_j\,+ \, \text{h.c.},  \nonumber  \\
& \mathcal{L}_{G^\pm}\, =\, -\frac{g_w}{\sqrt{2} M_W} \, G^- \,
\sum_{l=1}^{3}\sum_{j=1}^{3 + n_S} {\bf U}_{lj}   \bar \ell_l \left(
m_i P_L - m_j P_R \right) \nu_j\, + \, \text{h.c.}\,. 
\end{align}
where $P_{L,R} = (1 \mp \gamma_5)/2$. As is clear from the above equations,
flavour is violated by mixings in both charged and neutral current interactions. 
Denoting by $l = 1, \dots, 3$ the flavour of the charged leptons,
and by $i, j = 1, \dots, 3+n_S$ the physical neutrino states, the
mixing in charged current interactions is parametrized by a
rectangular $3 \times (3 +n_S)$ mixing matrix, ${\bf U}_{lj}$.
Notice that in the case of three neutrino generations, and
assuming alignment of the charged lepton's weak and 
mass basis,  
${\bf U}$ corresponds to the (unitary) PMNS matrix, $U_\text{PMNS}$.
For $n_\nu >3$ ($n_S \geq 1$), the 
mixing between the left-handed leptons, which we will subsequently 
denote by $\tilde U_\text{PMNS}$, 
corresponds to a $3 \times 3$ block of ${\bf U}$. 
One can parametrize the 
$\tilde U_\text{PMNS}$ mixing matrix as~\cite{FernandezMartinez:2007ms}
\begin{equation}\label{eq:U:eta:PMNS2}
U_\text{PMNS} \, \to \, \tilde U_\text{PMNS} \, = \,(\mathbb{1} - \eta)\, 
U_\text{PMNS}\,,
\end{equation}
where the  matrix $\eta$ encodes the deviation of $\tilde
U_\text{PMNS}$ from unitarity~\cite{Schechter:1980gr,Gronau:1984ct}, 
due to the presence of extra fermion states.
It is also convenient to introduce the invariant quantity
$\tilde \eta$, defined as
\begin{equation}\label{eq:def:etatilde}
\tilde \eta = 1 - |\text{Det}(\tilde U_\text{PMNS})| \, ,
\end{equation}
particularly useful to illustrate the effect of the new
active-sterile mixings (corresponding to a deviation from unitarity of
the $\tilde U_\text{PMNS}$).

As can be seen from above, the mixing in the neutral lepton sector
induced by the Majorana states 
also opens the possibility for flavour violation in neutral currents;
this is encoded in a square $ (3 +n_S) \times (3 +n_S)$ mixing matrix
\begin{equation}\label{eq:Cmatrix:def}
{\bf C}_{ij} \,=\,\sum_{l=1}^{3} {\bf U}_{l i}^*\,{\bf U}_{l j}\,. 
\end{equation}

\subsection{Rare lepton flavor violating $\boldsymbol{Z}$ decays
  revisited}\label{sec:LFVZdec}
\begin{figure}[!t]
\begin{center}
\begin{tabular}{ccc}
\centering
\includegraphics[width=40mm]{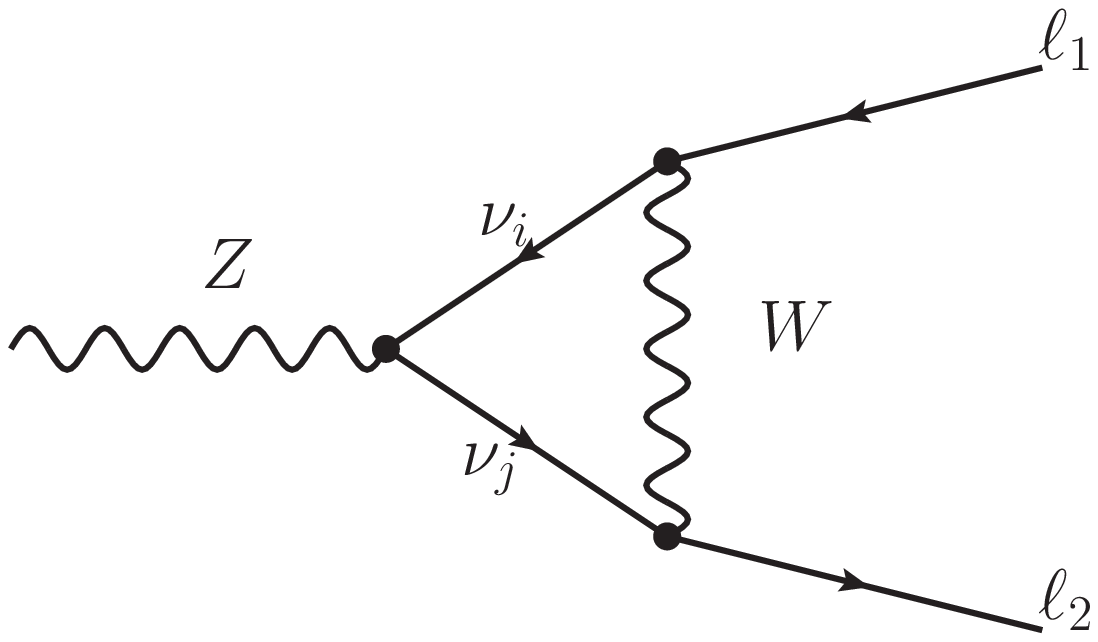}
&
\includegraphics[width=40mm]{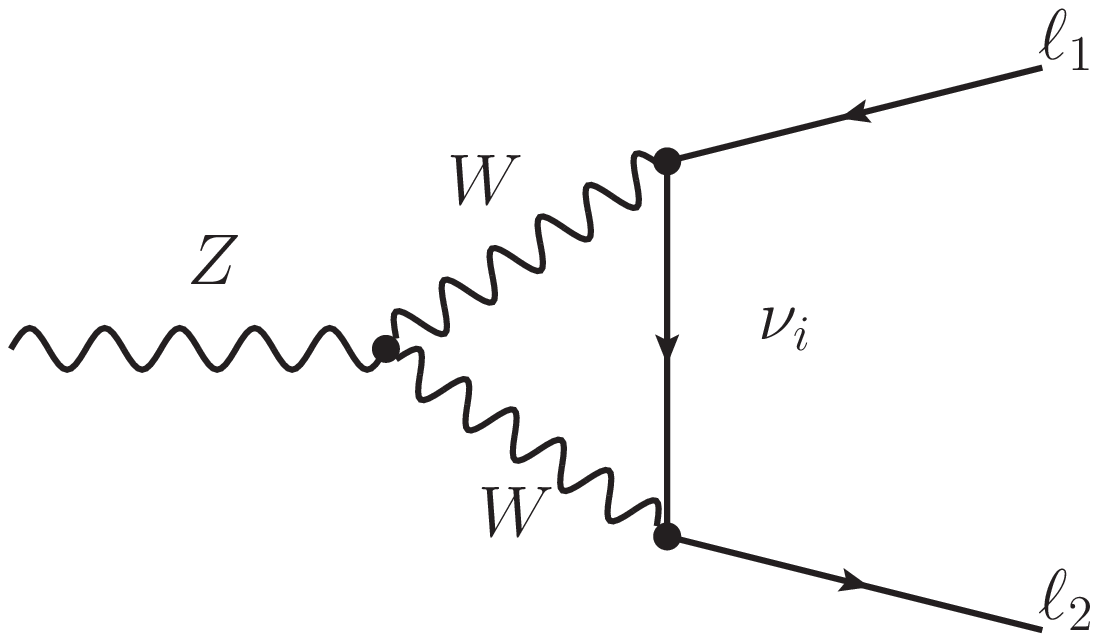}
&
\includegraphics[width=40mm]{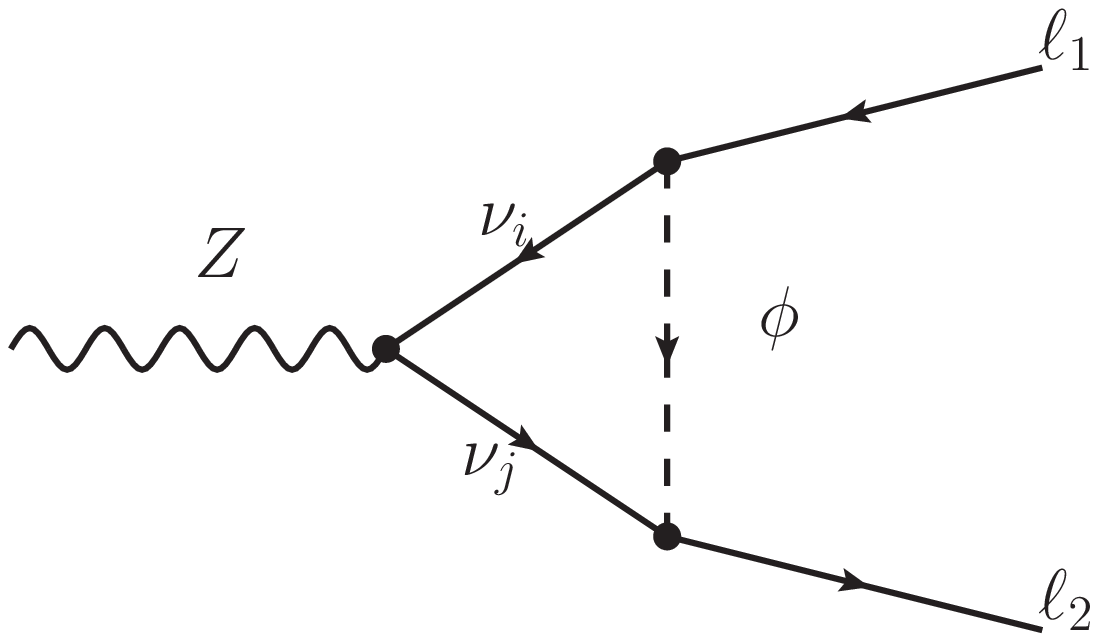}\\
\includegraphics[width=40mm]{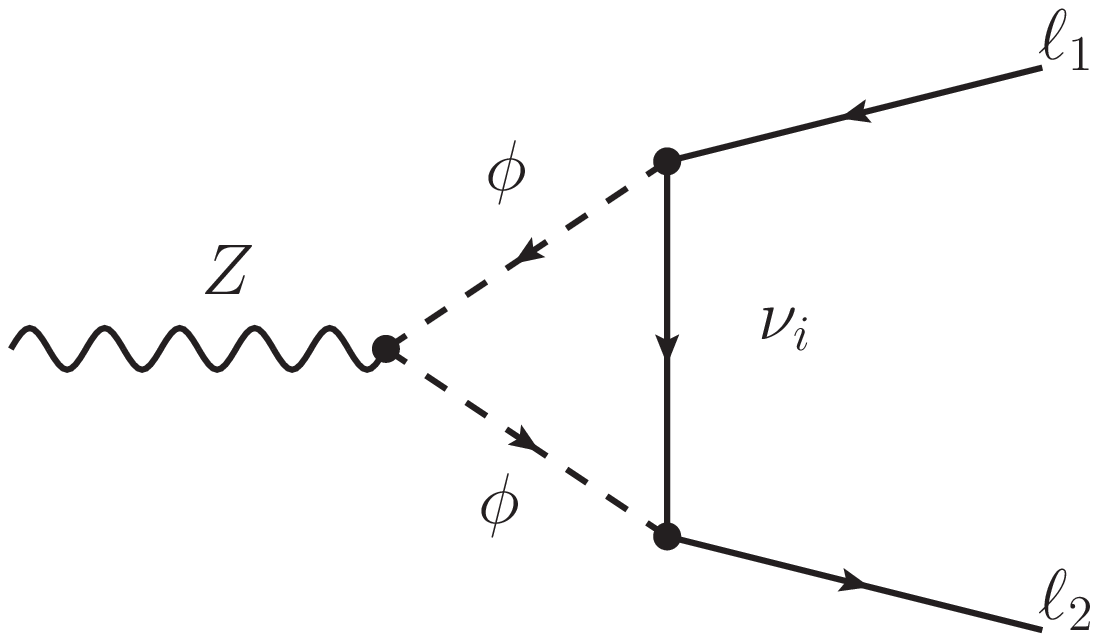}
&
\includegraphics[width=40mm]{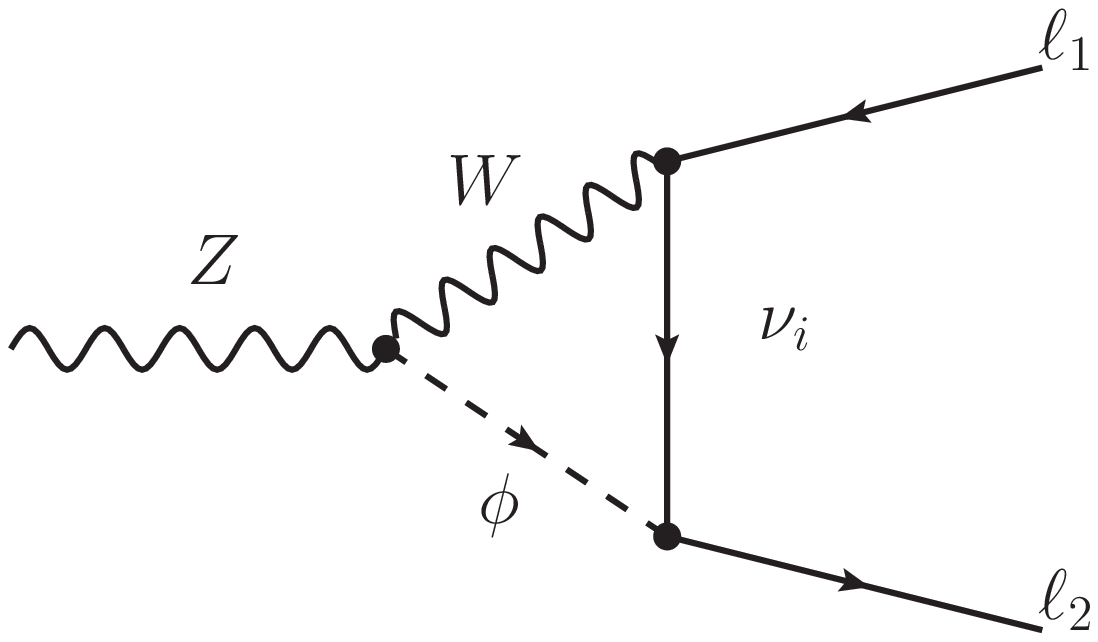}
& \raisebox{8mm}{ + crossed} \\
\end{tabular}
\begin{tabular}{cccc}
\centering
\includegraphics[width=35mm]{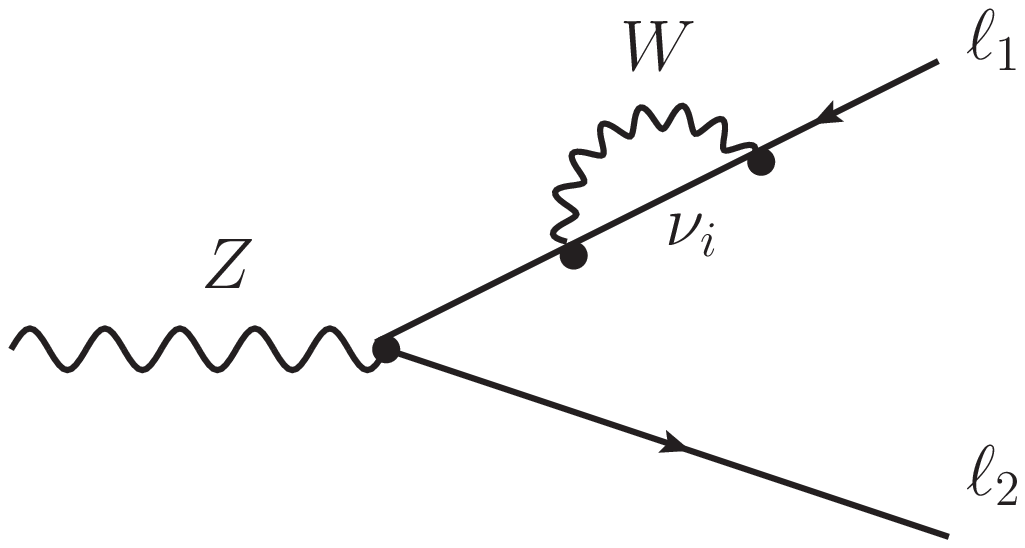}
&
\includegraphics[width=35mm]{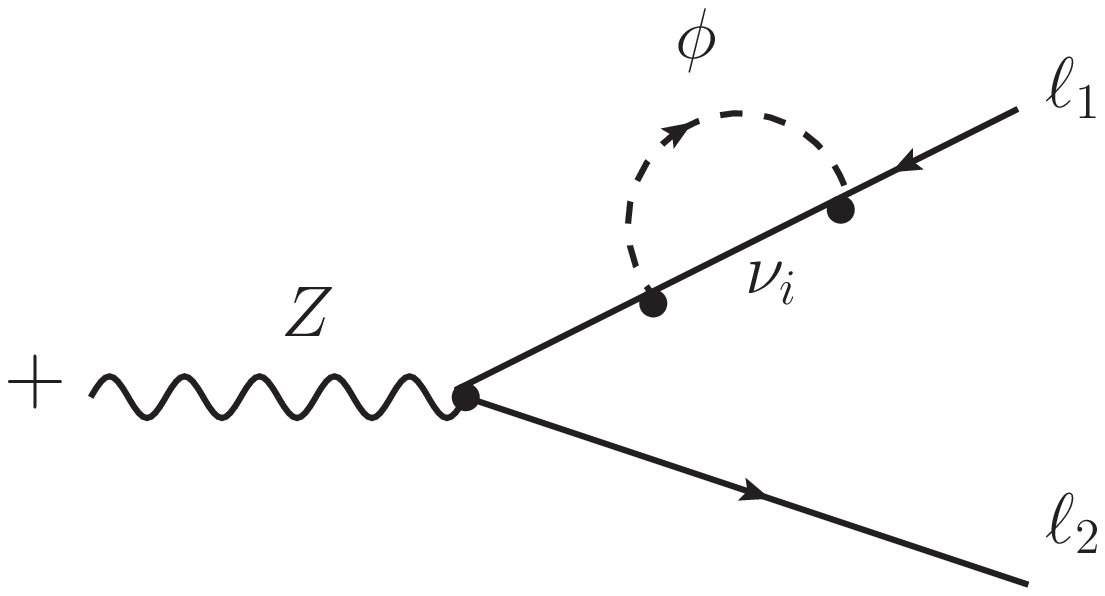}
&
\raisebox{-3mm}{\includegraphics[width=35mm]{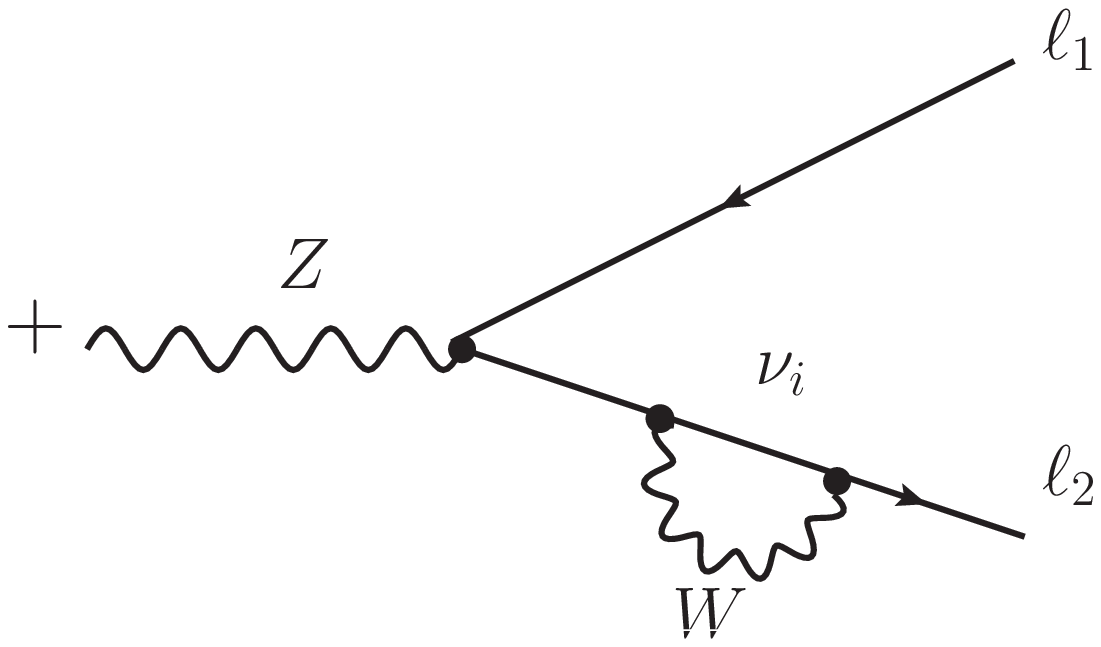}}
&
\raisebox{-3mm}{\includegraphics[width=35mm]{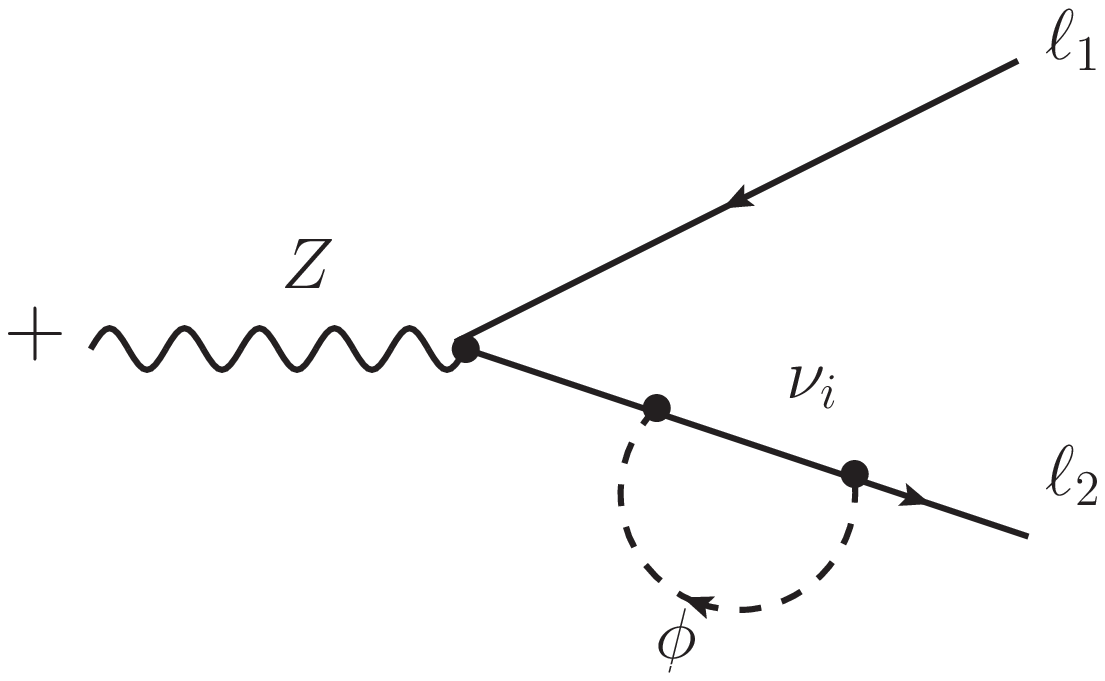}}
\end{tabular}
\end{center}
\caption{Feynman diagrams for the charged lepton-flavour changing $Z$
  decay. From left to right, top to bottom: $v_{W\nu \nu}(i,j),
  v_{WW\nu}(i) , v_{\phi \nu \nu}(i,j), v_{\phi\phi \nu}(i), 
  v_{W\phi \nu}(i)$. The last row contains
  the self-energy corrections to the external fermion legs, $v_{\rm
  SelfE}(i)$. 
}\label{fig:FeynDiag} 
\end{figure}

One of the main features of the SM extended by sterile Majorana neutrinos,
which mix with the active ones, is thus the possibility of flavour
violating $Z\nu_i\nu_j$ interactions (flavour-changing neutral
currents), coupling both the left- and right- handed components of
the neutral fermions to the $Z$ boson. Together with the
charged-current LFV couplings ($\propto \tilde U_\text{PMNS}$), 
these interactions will induce an 
effective {\it charged} lepton-flavour violating
vertex $Z \ell_1^\mp \ell_2^\pm$. 
We depict the full set of one-loop diagrams in Fig.~\ref{fig:FeynDiag}.  

Taking into account the contributions of all above higher order
processes, the branching ratio for cLFV $Z$ decays
(cf. Eq.~(\ref{eq:BR_LFVZ})) 
is given
by~\cite{Ilakovac:1994kj,Illana:1999ww,Illana:2000ic,Perez:2003ad,FloresTlalpa:2001sp,Delepine:2001di}:  
\begin{equation}\label{eq:BR_form_factor}
{\rm BR}(Z\to \ell_1^{\mp} \ell_2^{\pm})
\,=\, \frac{\alpha_W^3}{192\pi^2 \rm c_W^2}\frac{M_Z}{\Gamma_Z}
~|{\cal F}(M^2_Z)|^2
\approx 10^{-6}~|{\cal F}(M^2_Z)|^2, \quad \rm with ~\ell_1 \neq \ell_2.
\end{equation}

The form factor ${\cal F}(Q^2)$ encodes the details of the new interaction
and therefore the contribution of the sterile neutrinos: 
\begin{equation}\label{eq:FormFactor}
 {\cal F}(Q^2) \,=\, \sum_{i,j=1}^{n_\nu} {\bf U}_{l_1i} {\bf U}_{l_2j}^* ~
V_Z(x_i,x_j,x_Q)\,,
\end{equation}
where  $V_Z(x_i,x_j,x_Q)$ is the
vertex function, fully describing the amplitude, and which depends
quadratically on the neutrino masses. 
In the previous expression,  
we have introduced the mass ratios\footnote{The negligible effect of the final state charged lepton masses is ignored for simplicity here.}
$x_i = m_{\nu_i}^2/M_W^2$ and
the virtuality of the $Z$ boson $x_Q = Q^2/M^2_W$ (i.e., 
$x_Z = M_Z^2/M_W^2$ when it is on-shell). 
In the 't Hooft-Feynman gauge, all diagrams of
Fig.~\ref{fig:FeynDiag}~\cite{Ilakovac:1994kj,Illana:1999ww,Illana:2000ic,Perez:2003ad,FloresTlalpa:2001sp,Delepine:2001di} contribute to the amplitude $V_Z$:
\begin{equation}\label{l24}
V_Z(x_i,x_j,x_Q) \,=\,
v_{W\nu \nu}(i,j)+v_{WW\nu}(i)+v_{\phi \nu \nu}(i,j)+v_{\phi\phi
  \nu}(i)+v_{W\phi \nu}(i)+v_{\rm SelfE}(i)\,, 
\end{equation}
with the different contributions given in terms of 
dimensionless one-loop tensor integrals\footnote{The 
tensor integrals are numerically
evaluated via {\tt LoopTools}~\cite{Hahn:1998yk}, 
based on the {\tt  FF}~\cite{vanOldenborgh:1990yc} package, which
are linked to a private fortran code.} 
$\c_0$, $\cbar_0$, $\c_{ab}$, $\cbar_{ab}$ and
$\b_1$~\cite{'tHooft:1972fi,Passarino:1979jh},  
listed in Appendix~\ref{loops},
\begin{align}\label{D1}
v_{W\nu \nu}(i,j) & = \, 
-{\bf C}_{ij}\big[ x_Q(\c_{0}+\c_{11}+\c_{12}+\c_{23}) 
-2\c_{24}+1\big] + {\bf C}^*_{ij}\sqrt{x_ix_j}\ \c_0\,, \\ 
\label{D2}
v_{WW\nu}(i)  & = \, 2c^2_W\ (2I^{i_L}_3) \left[ x_Q\ 
({\cbar}_{11}+{\cbar}_{12}+{\cbar}_{23})-6{\cbar}_{24}+1\right]\,, \\  
\label{D3}
 v_{\phi \nu \nu}(i,j)  & = \, -{\bf C}_{ij}\displaystyle\frac{x_ix_j}{2}\c_0 
+ {\bf C}^*_{ij}\displaystyle\frac{\sqrt{x_ix_j}}{2}
 \left[x_Q\c_{23}-2\c_{24}+\displaystyle\frac{1}{2}\right]\,,\\
\label{D4}
v_{\phi\phi\nu}(i)  & = \,-(1-2s^2_W)\ (2I^{i_L}_3)\ x_i\ {\cbar}_{24}\,,\\  
\label{D5}
v_{W\phi\nu}(i)  & = \,-2s^2_W\ (2I^{i_L}_3)\ x_i\ {\cbar}_0 \,,\\
\label{Dse}
v_{\rm SelfE}(i)  & = \, 
\displaystyle\frac{1}{2}(v_i+a_i-4c^2_W a_i)\left[(2+x_i)\b_1 +1 \right]\,.
\end{align}
In the above, the weak neutral vector and axial-vector couplings are defined as
\begin{align}
\label{def-v}
v_i &=\, I^{i_L}_3 - 2 Q_i s_W^2 
\,,\\
\label{def-a}
a_i &= I^{i_L}_3\,,
\end{align}
with $s_W\, (c_W)$ denoting $\sin \theta_W$ ($\cos \theta_W$),  $Q_i$
the electric charge and $I^{i_L}_3$ the third component of weak
isospin.

\subsection{Lepton flavour universality in $\boldsymbol{Z}$
  decays}\label{sec:VLFUZdec} 
As mentioned before, in the SM the charged lepton couplings to the $Z$
boson are strictly flavour universal. Due to the $\tau$ lepton mass, 
$\Gamma (Z \to \tau^+ \tau^-)$ slightly differs from  $\Gamma (Z \to
\ell^+ \ell^-)$, with $\ell = e, \mu$ (see, e.g.~\cite{Freitas:2014hra}) 
\begin{equation}
\Gamma (Z \to \ell^+ \ell^-)\,=\,0.08397\,, \quad \quad  
\Gamma (Z \to \tau^+ \tau^-)\,=\,0.08378\,.
\end{equation}
The current experimental bound from LEP regarding (non-)universality
of $Z$ decays into electrons and muons is~\cite{Agashe:2014kda}
\begin{equation}
\frac{\Gamma^{\mu \mu, \rm SM}_Z}{\Gamma^{e e, \rm SM}_Z} = 1.0009 \pm
2.8 \times 10^{-3}\,. 
\end{equation} 
Assuming that the electron- and muon- partial widths are equal
($\Gamma^{e e, \rm SM}_Z = \Gamma^{\mu \mu, \rm SM}_Z$)
we can define the following observable
\begin{equation}
\label{eq:DeltaZ}
\Delta R^{\rm lep}_Z = 1 - \frac{\left (1 + \frac{\Gamma^{\mu \mu, \rm
      NP}_Z}{\Gamma^{\mu \mu, \rm SM}_Z} \right )}{\left (1+
  \frac{\Gamma^{e e, \rm NP}_Z}{\Gamma^{e e, \rm SM}_Z} \right )}\,,
\end{equation}
where $\Gamma^{\ell \ell, \rm NP}_Z$ refers to the contribution
induced by the sterile neutrinos.  

In our study, we will also investigate the contributions of the
sterile states to the width $\Gamma^{\ell \ell, \rm NP}_Z$, or
equivalently to the 
flavour conserving BR($Z \to \ell^+ \ell^-$), 
given by the diagonal contribution of Eq.~(\ref{eq:BR_form_factor}), in order to address the possibility of violation 
of lepton flavour universality (LFU).

\subsection{A high-luminosity $\boldsymbol{Z}$-factory}\label{sec:FCCee.exp}

Following the first evidence for a new (SM-like Higgs) 
bosonic resonance with a relatively low mass, the case for a
high luminosity circular $e^+e^-$ collider, operating at
centre-of-mass energies ranging from the $Z$ pole up to the top quark
pair threshold is being actively
studied~\cite{Gomez-Ceballos:2013zzn}. These initial investigations
are serving as a starting basis for a four-year design study of a $\sim
100$ km circumference $e^+e^-$ collider, which defines the framework
of the experimental prospects envisaged in this work. The baseline
design of this machine assumes a layout similar to LEP/LHC with a
number of equal-length arcs and long straight sections, in which the
two beams must circulate in separate vacuum chambers, leading to
${\cal O} (10^4)$ bunches for an operation at the $Z$ pole. These
characteristics should allow to obtain a typical peak luminosity at
the $Z$ pole of $\sim 10^{36} \rm cm^{-2} \rm s^{-1}$. A year of operation at the $Z$ pole
centre-of-mass energy would then yield  $\sim 10^{12}$ $Z$ boson
decays to be recorded. An alternative scheme, referred to as
``crab-waist'' scheme, could further increase the number of $Z$ decays by
an order of magnitude.  

The LFV $Z$ decays under scrutiny in this work imply a priori very
clean experimental signatures. For instance, the decay $Z \to
e^{\pm}\mu^{\mp}$  exhibits two and only two back-to-back
oppositely charged leptons originating from a unique vertex. The
decays $Z \to e^{\pm}(\mu^{\pm}) \tau^{\mp}$ could lead to somewhat
more ambiguous final states, depending on the subsequent $\tau$
decays. They can actually proceed leptonically (${\rm BR} (\tau \to
\ell \nu_{\ell}) \sim 17.5 \% $) or hadronically, being dominated in
the latter case by one- or three-prong decays. The direction of the
$\tau$ particle is given by the momentum of the opposite lepton and
hence can be used to kinematically constrain the decay. At least,
experimental studies with a realistic detector simulation and the
consideration of the relevant backgrounds are required to estimate the
performance of the reconstruction of the decays involving $\tau$
leptons.  In the following, we assume that the experimental reach for
these LFV decays is fully driven by the accessible luminosity. We
consider two bounds for the sensitivity: one $\sim {\cal O} (10^{-9})$
inspired by previous prospective studies at a Giga $Z$
factory~\cite{wilson} (or for a Linear Collider) and another $\sim
{\cal O} (10^{-13})$ corresponding to the highest foreseen luminosity
scheme ($10^{13}$ $Z$).

The parameter space of the models considered in this work is
constrained in particular by the present electroweak precision
measurements at the $Z$ pole. The unprecedented statistics which could
be obtained at FCC-ee are expected to improve significantly the
determination of some of these key constraints. 
Other expected precision improvements concern observables also used in
this work as is the case of the ratio of the partial widths of the $Z$
decays into electrons and muons. The current precision on this ratio is
at the level of  $2.8 \times 10^{-3}$~\cite{ALEPH:2005ab} and could be
increased by two orders of magnitude, ${\cal O} (5 
\times 10^{-5})$~\cite{Gomez-Ceballos:2013zzn}. 
The uncertainty of the partial
decay width of $Z \to \tau^+ \tau^-$ must accordingly decrease.  
Moreover, the
uncertainty on the invisible $Z$ width (expressed as the
number of light active neutrinos $N_{\nu}$) should also decrease from 0.008
to 0.00004, by only a scaling of the uncertainty with the expected
statistics. Nevertheless, 
the main systematic limitation comes from the luminosity
measurement and must be accordingly evaluated. A reasonable target for the
uncertainty on the number of neutrinos at FCC-ee is estimated at
${\cal O} (0.001)$~\cite{Gomez-Ceballos:2013zzn}.  

\section{Constraints on sterile neutrino extensions of the
  SM}\label{sec:constraints}

In order to account for neutrino masses and mixings, many extensions
of the SM call upon the introduction of RH neutrinos (giving
rise to a Dirac mass term for the neutral leptons) and/or other new
particles.  Their phenomenological impact can be important if 
the sterile states are not excessively heavy, and have 
sizeable mixings to the light (mostly active) neutrinos. 
For instance, this is the case 
of the $\nu$MSM~\cite{Asaka:2005an}, the Inverse Seesaw
(ISS)~\cite{Mohapatra:1986bd} and the low-scale
type-I seesaw~\cite{Ibarra:2010xw}. 
Many observables will be sensitive to the 
active-sterile mixings, and their current experimental values (or
bounds) will thus constrain such SM extensions. 
In what follows we proceed to discuss the most relevant constraints on
models with sterile fermions. 

\bigskip
\noindent{\bf Neutrino oscillation data}

\noindent
The most important constraint on any model of massive
neutrinos is to comply with $\nu$-oscillation 
data~\cite{Tortola:2012te,Fogli:2012ua,GonzalezGarcia:2012sz,Forero:2014bxa,nufit,Gonzalez-Garcia:2014bfa}.
In our analysis, we 
consider both normal and inverted hierarchies for the 
light neutrino spectrum~\cite{Forero:2014bxa}; the corresponding
best-fit intervals in the case of normal hierarchy (NH) are
\begin{align}\label{eq:neutrino.data.NH}
\sin^2 \theta_{12}\,=\, 0.323\,, 
\quad
\sin^2 \theta_{23}\,=\, 0.567 \,, 
\quad
\sin^2 \theta_{13}\,=\, 0.0234\,, \nonumber \\
\Delta m^2_{21}\,=\, 7.60\times 10^{-5} \mathrm{eV}^2\,, 
\quad
|\Delta m^2_{31}|\,=\, 2.48\times 10^{-3} \mathrm{eV}^2\, ,
\end{align}
whereas for an inverted mass hierarchy (IH) the values are
\begin{align}\label{eq:neutrino.data.IH}
\sin^2 \theta_{12}\,=\, 0.323\,, 
\quad
\sin^2 \theta_{23}\,=\, 0.573\,, 
\quad
\sin^2 \theta_{13}\,=\, 0.024\,, \nonumber \\
\Delta m^2_{21}\,=\, 7.60\times 10^{-5} \mathrm{eV}^2\,, 
\quad
|\Delta m^2_{31}|\,=\, 2.38\times 10^{-3} \mathrm{eV}^2\, . 
\end{align}

The value of the CP violating Dirac phase $\delta$ is still
undetermined, although the complementarity of accelerator and reactor
neutrino data starts reflecting in a better sensitivity to the CP
violating phase $\delta$ \cite{Capozzi:2013csa,Forero:2014bxa} (and to 
the hierarchy of the light neutrino spectrum).

\bigskip
\noindent{\bf Unitarity constraints}

\noindent
The introduction of fermionic sterile states can give
rise to non-standard neutrino interactions with matter.  
Bounds on the non-unitarity matrix $\eta$ (cf. Eq.(\ref{eq:U:eta:PMNS2})),
have been derived in~\cite{Antusch:2008tz,Antusch:2014woa}  
by means of an effective theory approach.  
We apply them in our numerical analysis for the cases in which the
latter approach is valid, generically for sterile masses above the GeV, but 
below the electroweak scale, $\Lambda_\text{EW}$.

\bigskip
\noindent{\bf Electroweak precision data}

\noindent
Electroweak (EW) precision constraints on sterile fermions were
firstly addressed in~\cite{delAguila:2008pw} with an effective
approach (and therefore valid only for multi-TeV singlet states). 
The impact of sterile neutrinos on the invisible $Z$-decay width has also been 
addressed in~\cite{Akhmedov:2013hec,Basso:2013jka,Abada:2013aba}, where it
has been shown that $\Gamma(Z \to \nu \nu)$ can be reduced 
with respect to the SM prediction. 
Indeed, the addition of sterile states to the SM with a sizeable active-sterile
mixing may have an impact on 
the electroweak precision observables either at
tree-level (charged currents) or at higher order. 
In particular, the non-unitarity of the active neutrino
mixing matrix, Eq.~(\ref{eq:U:eta:PMNS2}), implies that the 
couplings of the active neutrinos to the $Z$ and $W$
bosons are suppressed with respect to their SM values. 
Complying with  LEP results
on $\Gamma(Z \to \nu \nu)$~\cite{Agashe:2014kda} will then also constrain
these sterile neutrino extensions.
In addition, we further require that the new contributions to the LFV
$Z$ decay width do not exceed the present uncertainty on the 
total $Z$ width~\cite{Agashe:2014kda}:
$\Gamma (Z \to \ell_1^\mp \ell_2^\pm) < 
\delta \Gamma_{\rm  tot}$.  

\bigskip
\noindent{\bf LHC constraints}

\noindent
The presence of a new Higgs boson decay channel 
with (heavy) neutrinos in the final state can enlarge the total
Higgs decay width, thus lowering the SM predicted decay branching
ratios. LHC data already allows to constrain
regimes where the sterile states are below the Higgs mass, 
due to the potential Higgs decays to an active and heavier (mostly) sterile
neutrinos. In our analysis we apply the constraints derived 
in~\cite{BhupalDev:2012zg,Cely:2012bz,Bandyopadhyay:2012px}.

\bigskip
\noindent{\bf Leptonic and semileptonic meson decays}

\noindent
Further constraints arise from leptonic and semileptonic decays of
pseudoscalar mesons 
$\ K,\ D, \ D_s$, $B$  (see~\cite{Goudzovski:2011tc,Lazzeroni:2012cx} for
kaon decays,~\cite{Naik:2009tk,Li:2011nij} for $D$ and $D_S$ decay
rates, and~\cite{Aubert:2007xj,Adachi:2012mm} for $B$-meson observations).
These decays have been addressed in~\cite{Abada:2012mc,Abada:2013aba}
in the framework of the SM extended by sterile neutrinos, and it was found that 
the most severe bounds arise from the violation of lepton universality
in leptonic kaon 
decays (parametrized by the observable 
$\Delta r_K$), which can receive important contributions from
the new sterile states, due to the new phase space factors, 
and as a result of deviations from unitarity of the $\tilde
U_\text{PMNS}$. 

\bigskip
\noindent{\bf Laboratory searches}

\noindent Negative searches for monochromatic lines in the
spectrum of muons from  $\pi^\pm \to \mu^\pm \nu$
decays~\cite{Kusenko:2009up,Atre:2009rg} also impose robust bounds on 
sterile neutrino masses in the MeV-GeV range. 

\bigskip
\noindent{\bf Lepton flavour violation}

\noindent
Depending on the sterile neutrino mass regime, and on the 
active-sterile mixings, the new states will contribute to several 
charged lepton flavour violating processes such as  
$\ell \to \ell^\prime \gamma$, $\ell\to \ell_1\ell_1\ell_2$ and
$\mu-e$ conversion in muonic atoms.  
In our analysis we compute the contribution of the sterile states to
all these
observables~\cite{Ma:1979px,Gronau:1984ct,Ilakovac:1994kj,Deppisch:2004fa,Deppisch:2005zm,Dinh:2012bp,Alonso:2012ji,Abada:2014kba}, 
imposing compatibility with the bounds summarised in
Table~\ref{table:cLFV:bounds}, 
also considering the impact of the future experimental sensitivities.
\begin{table}[tb!]
\centering
\begin{tabular}{|c|c|c|}
\hline
cLFV Process & Present Bound & Future Sensitivity  \\
\hline
    $\mu \rightarrow  e \gamma$ & $5.7\times
10^{-13}$~\cite{Adam:2013mnn}  & $6\times
10^{-14}$~\cite{Baldini:2013ke} \\ 
    $\tau \to e \gamma$ & $3.3 \times 10^{-8}$~\cite{Aubert:2009ag}& $
\sim3\times10^{-9}$~\cite{Aushev:2010bq}\\ 
    $\tau \to \mu \gamma$ & $4.4 \times 10^{-8}$~\cite{Aubert:2009ag}&
$ \sim3\times10^{-9}$~\cite{Aushev:2010bq} \\ 
    $\mu \rightarrow e e e$ &  $1.0 \times
10^{-12}$~\cite{Bellgardt:1987du} &
$\sim10^{-16}$~\cite{Blondel:2013ia} \\ 
    $\tau \rightarrow \mu \mu \mu$ &
$2.1\times10^{-8}$~\cite{Hayasaka:2010np} & $\sim
10^{-9}$~\cite{Aushev:2010bq} \\ 
    $\tau \rightarrow e e e$ &
$2.7\times10^{-8}$~\cite{Hayasaka:2010np} &  $\sim
10^{-9}$~\cite{Aushev:2010bq} \\ 
    $\mu^-, \mathrm{Ti} \rightarrow e^-, \mathrm{Ti}$ &  $4.3\times
10^{-12}$~\cite{Dohmen:1993mp} & $\sim10^{-18}$~\cite{Alekou:2013eta}\\ 
    $\mu^-, \mathrm{Au} \rightarrow e^-, \mathrm{Au}$ & $7\times
10^{-13}$~\cite{Bertl:2006up} & \\ 
    $\mu^-, \mathrm{Al} \rightarrow e^-, \mathrm{Al}$ &  &
$10^{-15}-10^{-18}$~\cite{Kuno:2013mha} \\ 
\hline
\end{tabular}
\caption{Current experimental bounds and future sensitivities for the
low-energy cLFV observables considered in our study.}
\label{table:cLFV:bounds}
\end{table}

\bigskip
\noindent{\bf Neutrinoless double beta decay}

\noindent
The introduction of singlet neutrinos with Majorana masses allows for 
new processes like lepton number violating interactions, among which 
neutrinoless double beta decay remains the most important
one \cite{Benes:2005hn}. 
In the SM extended by $n_S$ sterile states, the effective neutrino
mass $m_{ee}$ is given 
by~\cite{Blennow:2010th}: 
\begin{equation} \label{eq:22bbdecay}
 m_{ee}\,\simeq \,\sum_{i=1}^{3+n_S} {\bf U}_{ei}^2 \,p^2
\frac{m_i}{p^2-m_i^2} \simeq 
\left(\sum_{i=1}^3 {\bf U}_{ei}^2\, m_{\nu_i}\right)\, 
+ p^2 \, \left(\sum_{i=4}^{3+n_S} 
{\bf U}_{ei}^2 \,\frac{m_i}{p^2-m_i^2}\right)\,,
\end{equation}
where $p^2 \simeq - (100 \mbox{ MeV})^2$ is an average estimate over
different values depending on  
the decaying nucleus of the  virtual momentum of
the neutrino.\\ 
The neutrinoless double beta decay process is being actively searched
for by several 
experiments, by means of the  best performing detector techniques: 
among others, GERDA~\cite{Agostini:2013mzu}, 
EXO-200~\cite{Auger:2012ar,Albert:2014awa},
KamLAND-ZEN~\cite{Gando:2012zm} have all set strong bounds on the  
effective mass, 
to which the amplitude of $0\nu 2 \beta$  process is proportional. 
The sensitivities of current experiments put a limit on the effective
neutrino Majorana mass - determining the 
amplitude of the neutrinoless double beta decay rate - in the range 
\beq
| m_{ee}| \lsim 140\text { meV} - 700\text { meV}\,.
\eeq
In Table~\ref{tab:nulesssensitivities}, we summarise the future 
sensitivity of ongoing and planned $0\nu 2 \beta$ experiments.

\begin{table}[h!]
\begin{center}
{\begin{tabular}{| l | l | c |}  \hline                       
Experiment & Ref. &  $ |m_{ee} | $ (eV) \\
  \hline                       
 EXO-200 (4 yr) & \cite{Auger:2012ar,Albert:2014awa} & 0.075 - 0.2  \\
nEXO (5 yr)  & \cite{DeliaTosionbehalfoftheEXO:2014zza}& 0.012 - 0.029  \\
nEXO (5 yr + 5 yr w/ Ba tagging) &
\cite{DeliaTosionbehalfoftheEXO:2014zza} 
& 0.005 - 0.011  \\
KamLAND-Zen (300~kg, 3 yr)& \cite{Gando:2012zm}  & 0.045 - 0.11 \\
GERDA  phase II & \cite{Agostini:2013mzu} & 0.09 - 0.29 \\
CUORE (5 yr) & \cite{Gorla:2012gd,Artusa:2014lgv} & 0.051 - 0.133 \\
SNO+  & \cite{Hartnell:2012qd} & 0.07 - 0.14 \\
SuperNEMO & \cite{Barabash:2012gc} & 0.05 - 0.15 \\
NEXT & \cite{Granena:2009it,Gomez-Cadenas:2013lta}& 0.03 - 0.1 \\
MAJORANA demo. & \cite{Wilkerson:2012ga} & 0.06 - 0.17 \\
 \hline                       
\end{tabular}
}
\caption{Future sensitivity of several $0\nu 2 \beta$ experiments.}
\label{tab:nulesssensitivities}
\end{center}
\end{table}

\noindent
In our analysis, we consider this observable using the most recent
constraint from~\cite{Albert:2014awa}; concerning the future
sensitivity we take
$|m_{ee}| \lesssim 0.01$ eV.

\bigskip
\noindent{\bf Cosmological bounds}

\noindent A number of
cosmological observations~\cite{Smirnov:2006bu,Kusenko:2009up}
 put severe constraints on sterile neutrinos with a mass below the TeV. 
 While CMB analysis with the Planck satellite disfavour very light
 sterile neutrinos (with a mass $\lesssim$ eV)~\cite{Ade:2013zuv},   
 a $\sim$keV sterile
neutrino may instead be a viable DM candidate, also
offering a possible
explanation for  the observed X-ray line in galaxy clusters spectra at an
energy $\sim 3.5$~keV~\cite{Bulbul:2014sua,Boyarsky:2014jta}  and for 
the origin of pulsar kicks, or even to the baryon asymmetry of the
Universe (for a review see~\cite{Abazajian:2012ys}). 

The cosmological bounds are in general derived by
assuming the minimal possible abundance (in agreement with neutrino
oscillations) of sterile neutrinos in halos consistent with standard
cosmology. However,  the possibility of a
non-standard cosmology with a very low reheating temperature or a
scenario where the sterile 
neutrinos couple to  a dark sector~\cite{Dasgupta:2013zpn}, could
allow to evade some of the above bounds,  as argued in~\cite{Gelmini:2008fq}.
In this analysis,  aiming at being
conservative,  we will allow for the violation
of these cosmological bounds in some scenarios, explicitly stating
it.

\section{A minimal ``3+1 toy model''}\label{sec:3+1}
The most simple approach to studying the phenomenological 
impact of sterile fermions lies in considering a minimal model,
where one extra sterile Majorana state
is added to the three light active neutrinos of the SM.

\subsection{The ``3+1'' framework}
In the present framework, no assumption is made on the underlying mechanism of
neutrino mass generation. In addition to the three (light) active
masses and corresponding mixing angles, it is only assumed 
that the leptonic sector contains extra degrees of
freedom: the mass of the new sterile state, $m_4$, three
active-sterile mixing angles $\theta_{i4}$, two new (Dirac) CP phases
and one extra Majorana phase. This leads to the definition of a $4 \times 4$ mixing matrix ${\bf U}_{ij}$,
whose $3 \times 4$ sub-matrix ${\bf U}_{lj}$ appears in Eq.~(\ref{eq:lagrangian:WGHZ}).

Although the experimental and observational
constraints mentioned in Section~\ref{sec:constraints} put no upper
limit on the mass of the heavy neutrino, 
we notice however that the decay of the (mostly) 
sterile heavy states should comply with the perturbative unitary
condition~\cite{Chanowitz:1978mv,Durand:1989zs,Korner:1992an,Bernabeu:1993up,Fajfer:1998px, 
Ilakovac:1999md},
\begin{equation}\label{eq:sterile:perturbativity:mass:width}
\frac{\Gamma_{\nu_i}}{m_{\nu_i}}\, < \, \frac{1}{2}\, \quad \quad (i \geq 4)\,.
\end{equation}
Assuming that the sterile mass is indeed sufficiently large to allow for
its 2-body decay into a $W^\pm$ boson and a charged lepton, or into a 
light (active) neutrino and either a $Z$ or a Higgs boson, the total decay  
width of such a state ($i \geq 4$) is given by 
\begin{equation}\label{eq:sterile:decaywidth}
\Gamma_{\nu_i}\, =\, \sum_{j=1}^3 \left[ \Gamma({\nu_i} \to \ell_j W) + 
\Gamma({\nu_i} \to \nu_j Z) + \Gamma({\nu_i} \to \nu_j H) \right]\,
\approx \,\frac{\alpha_w}{4\,M^2_W} {\bf C}_{ii}\,,
\end{equation}
where $\alpha_w=g^2_w/4 \pi$, and ${\bf C}_{ii}$ as given in
Eq.~(\ref{eq:Cmatrix:def}). 
Since the dominant contribution arises from the 
charged current term, one is led 
to the following bound on the sterile masses and
their couplings to the active 
states~\cite{Chanowitz:1978mv,Durand:1989zs,Korner:1992an,Bernabeu:1993up,Fajfer:1998px, 
Ilakovac:1999md}:
\begin{equation}\label{eq:sterile:bounds:Ciimi}
m_{\nu_i}^2\,{\bf C}_{ii} \, < 2 \, \frac{M^2_W}{\alpha_w}\, \quad
\quad (i \geq 4)\,. 
\end{equation}

In our analysis, and for both NH and IH light neutrino spectra, we
scan over the following range for the sterile neutrino mass
\begin{equation}\label{eq:effective:m4range}
10^{-9} \text{ GeV }\lesssim \, m_4 \, \lesssim 10^{6} \text{ GeV }\,,
\end{equation}
while the active-sterile mixing angles are randomly varied in the
interval $[0, 2 \pi]$, always ensuring that the condition
of Eq.~(\ref{eq:sterile:bounds:Ciimi}) is respected. 
All CP phases are also taken into account, and
likewise randomly varied between 0 and $2 \pi$.

\subsection{LFU violation: $Z \to \ell \ell$ decays in the 
  ``3+1 model'' }
We begin by addressing the contributions of the additional
sterile state to the violation of flavour universality, considering
the observable $\Delta R^\text{lep}_Z$, introduced in Eq.~(\ref{eq:DeltaZ}). 
Although one could have a 
non-negligible violation of LFU $\sim \mathcal{O}(10^{-3})$, a number of
experimental bounds (LFV constraints, complying with $U_\text{PMNS}$
data, ...) preclude this possibility, 
and one has at most $\Delta R^\text{lep}_Z \lesssim
10^{-10}$, clearly beyond experimental sensitivity.

\subsection{LFV $Z$ decays in the ``3+1 model''}
\begin{figure}
\begin{center}
\begin{tabular}{cc}
\epsfig{file=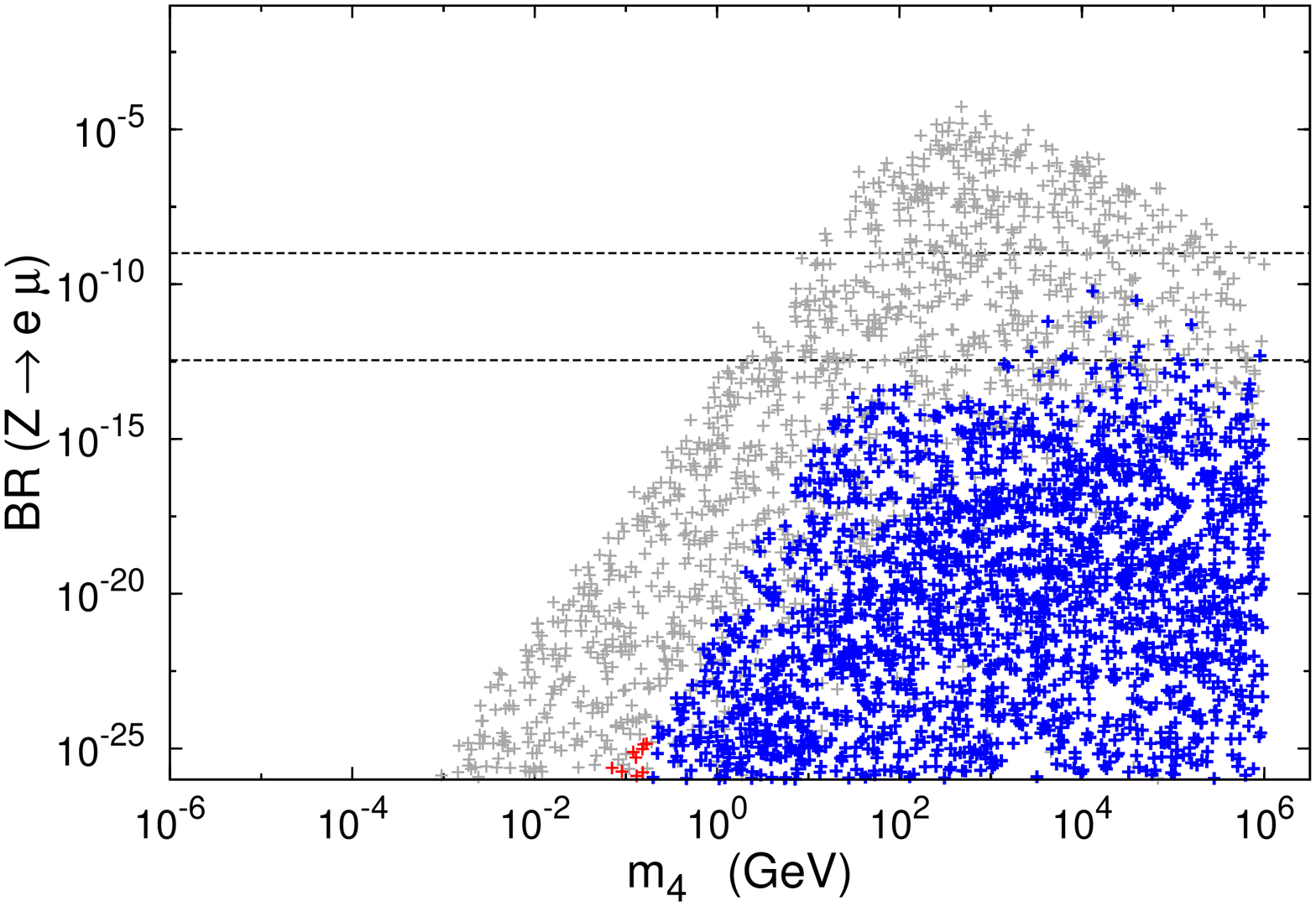, width=55mm,
  angle=270}
&
\epsfig{file=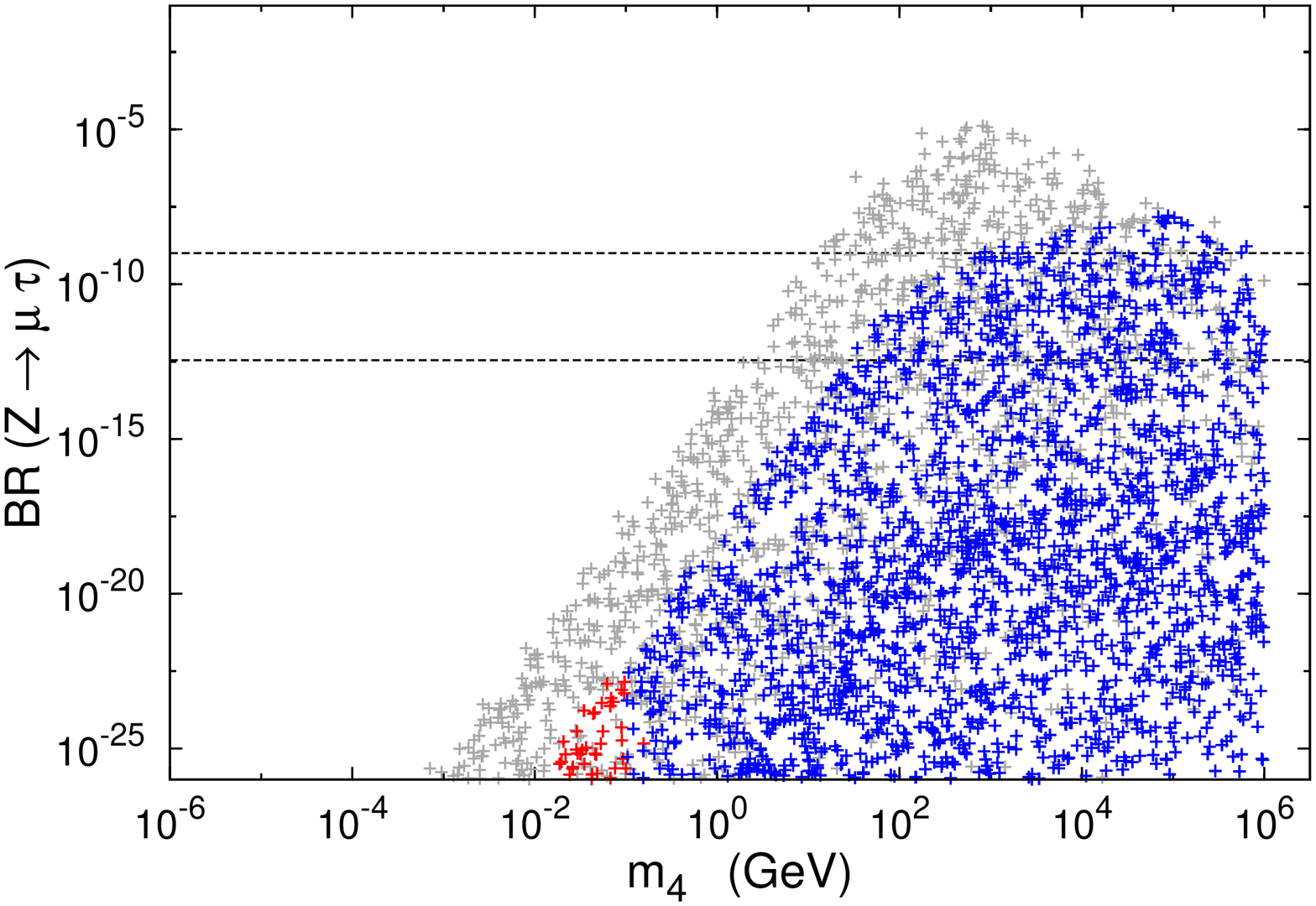, width=55mm, angle=270}
\end{tabular}
\end{center}
\caption{The ``3+1 model": on the left BR($Z \to e \mu$) and on the
  right BR($Z \to \mu \tau$), 
  as a function
  of the mass of the (mostly) sterile state, $m_4$, for a NH light
  neutrino spectrum. Blue
  points are in agreement with cosmological bounds, while the red ones
  would require considering a non-standard cosmology. In grey we
  denote points already excluded by other (non-cosmological) bounds
  (see text for a description). The upper horizontal dashed line 
  corresponds to the expected sensitivity for a GigaZ facility as a
  Linear Collider, $\mathcal{O}(10^{-9})$,
  the lower one to the FCC-ee $\sim \mathcal{O}(10^{-13})$. 
}\label{fig:3+1:BRmt.m4}
\end{figure}

We proceed to discuss the impact of the additional sterile
state regarding lepton flavour violating $Z$ decays. 
In Fig.~\ref{fig:3+1:BRmt.m4} we illustrate our results 
regarding the observation of BR($Z \to e \mu$) and BR($Z \to
\mu \tau$) at a future high-luminosity $Z$-factory, considering a NH
light neutrino spectrum (the results for an IH spectrum do not exhibit
any significant qualitative nor quantitative difference in what
concerns the branching fractions, and so we
will not display them here). As already mentioned in
Section~\ref{sec:constraints}, we identify in red the points that 
are typically disfavoured from standard cosmology arguments. 
Grey points denote failure to comply with (at least) one of the
following constraints: $\nu$-oscillation data, bounds on 
the $U_\text{PMNS}$ matrix,
bounds from EW precision data, LHC bounds,
laboratory bounds, constraints from
rare leptonic meson decays; conflict with bounds from 
cLFV decays, 
neutrinoless double beta decays or $Z$-boson decay width data
(invisible and lepton flavour conserving). Blue points are in
agreement with {\it all} imposed constraints.

As can be seen from Fig.~\ref{fig:3+1:BRmt.m4}, such a minimal
extension of the SM can indeed account\footnote{In addition to being
  experimentally ruled out, 
  we notice that very large branching fractions, associated with a
  regime of masses above the TeV, would be precluded due to the
  perturbativity bound of Eq.~(\ref{eq:sterile:bounds:Ciimi}), which
  significantly constraints the sterile-active mixings for heavy
  sterile states.} for values of BR($Z \to
\ell_1^\mp \ell_2^\pm$) within the sensitivity of a high luminosity
$Z$-factory, such as the FCC-ee. 
(We notice that we have only displayed here values of the (mostly)
sterile state mass $m_4 \gtrsim 10^{-3}$ GeV, since smaller values are
associated to BR($Z \to \ell_1^\mp \ell_2^\pm$)~$\lesssim 10^{-28}$).

Despite the potential of this simple ``toy-model'' to account for 
significant LFV $Z$ decay branching
fractions (which could be as large as $\mathcal{O}(10^{-6})$), these
cannot be reconciled with current bounds on 
low-energy cLFV processes (see Table~\ref{table:cLFV:bounds}), to
which the sterile states also contribute. While the recent MEG bound on 
$\mu \to e \gamma$ decays excludes important regions of the parameter
space\footnote{The flavour violating $Z \ell_1^\mp \ell_2^\pm$ vertex might
  induce higher order (2-loop) contributions to radiative muon
  decays~\cite{FloresTlalpa:2001sp}; however, in the present study, we
do not take such contributions into account.}, 
the contribution of the $Z$ penguin diagrams to cLFV 3-body decays and
$\mu-e$ conversion in nuclei severely
constrains the flavour violating $Z \ell_1^\mp \ell_2^\pm$ vertex 
(see also~\cite{Delepine:2001di,FloresTlalpa:2001sp,Perez:2003ad}).
This is especially manifest in the case of $Z \to e \mu$ decays, since
the severe constraints from BR($\mu \to 3 e$) and CR($\mu-e$, Au) typically
preclude BR$(Z \to e \mu) \gtrsim 10^{-13}$; however, and for a regime
of very heavy sterile states ($m_4 \gtrsim 10^4$~GeV), 
the ``3+1 model" can nevertheless account for 
BR($Z \to e \mu$) within FCC-ee reach.

The comparatively less stringent bounds for cLFV in the $\mu-\tau$
sector allow for larger BR($Z \to \mu \tau$): values 
above $\mathcal{O}(10^{-13})$ can be found for
$m_4 \gtrsim 50$~GeV, and even larger branching
fractions, $\mathcal{O}(10^{-8})$ (within reach of a GigaZ facility as a 
Linear Collider) for $m_4 \gtrsim 500$~GeV. 
Although not displayed here, the predictions of the ``3+1 model" for 
the BR($Z \to e \tau$) exhibit a similar behaviour to what is observed for
$Z \to \mu \tau$ decays.  

\begin{figure}
\begin{tabular}{cc}
\epsfig{file=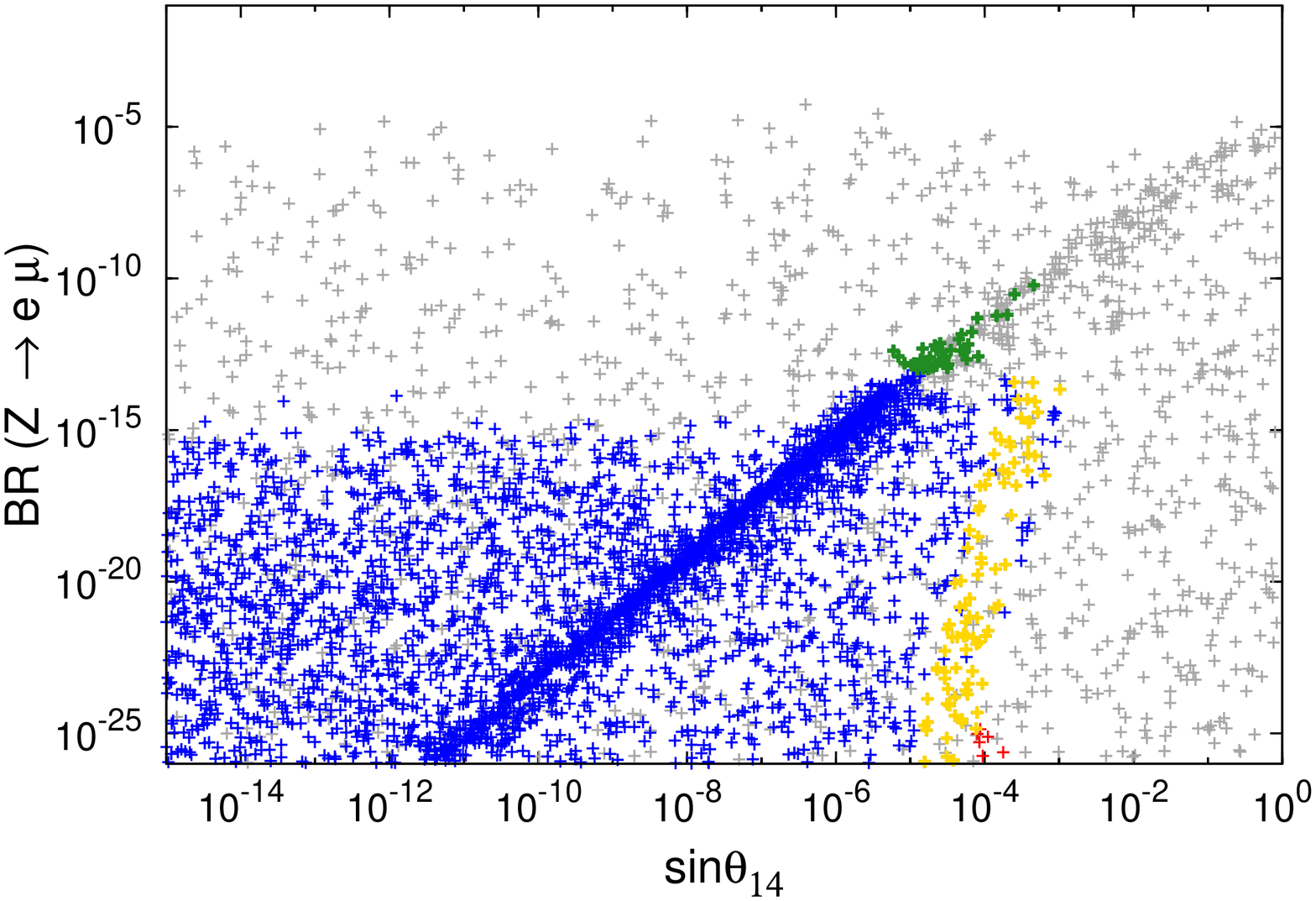, width=55mm, angle=270}
&
\epsfig{file=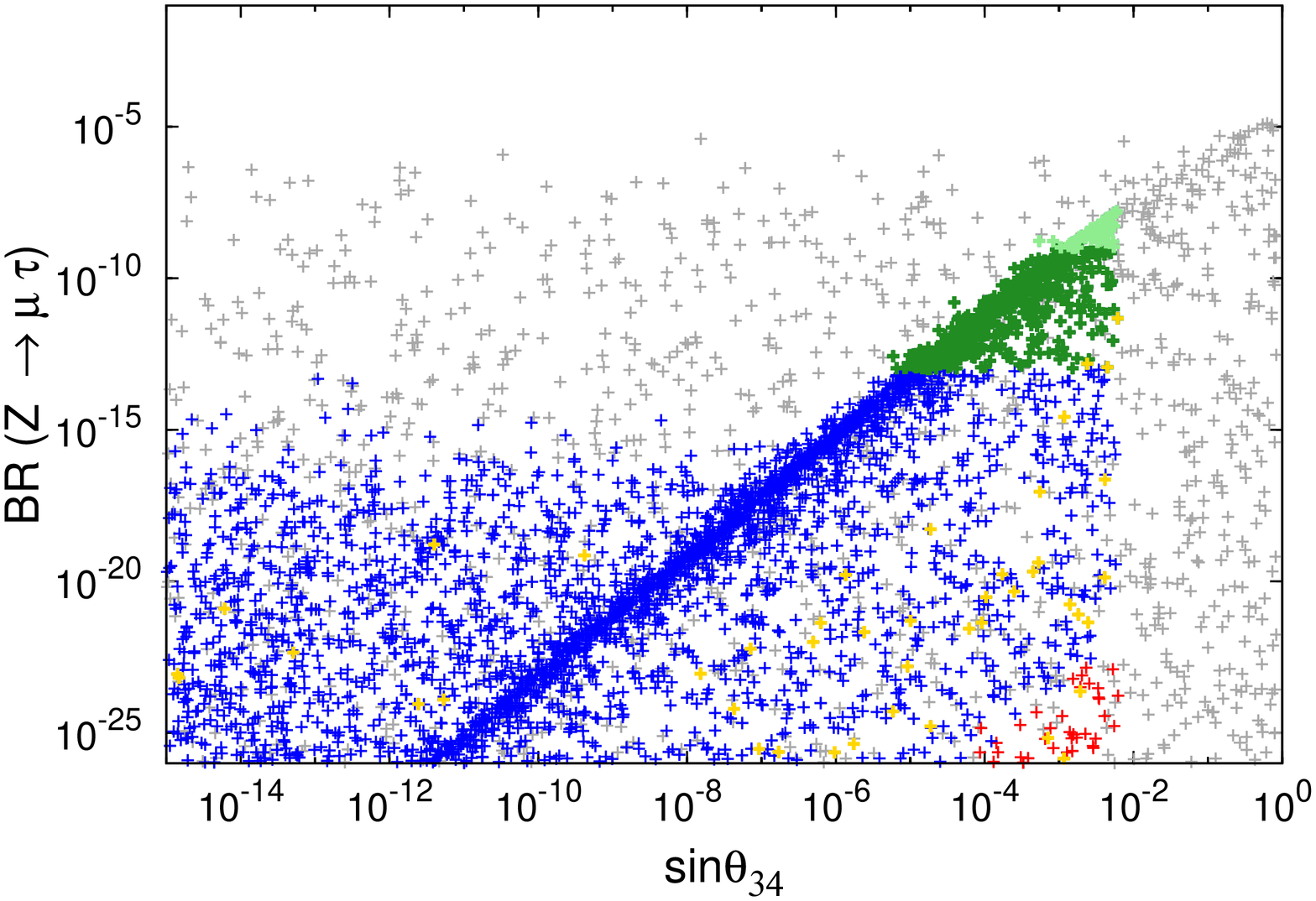, width=55mm, angle=270}
\end{tabular}
\caption{The ``3+1 model": BR($Z \to e \mu$) as a function of
  the active-sterile mixing $\theta_{14}$ (left) and 
  BR($Z \to \mu \tau$) as a function of $\theta_{34}$
  (right) for a NH light neutrino spectrum. Blue
  points are in agreement with cosmological bounds, while the red ones
  would require considering a non-standard cosmology. In grey we
  denote points already excluded by other (non-cosmological) bounds
  (see text for a description); dark-yellow
  points denote an associated $|m_{ee}|$ within
  experimental reach (i.e. $0.01 \text{ eV} \lesssim |m_{ee}|\lesssim 0.1
  \text{ eV}$). Dark green points are associated
  with $10^{-13} \lesssim $ BR($Z \to \ell_1^\mp \ell_2^\pm$) 
  $\lesssim 10^{-9} $,
  while light green ones correspond to 
  BR($Z \to \ell_1^\mp \ell_2^\pm$) $\gtrsim 10^{-9}$.
}\label{fig:3+1:BR.theta14.theta34}
\end{figure}

The r\^ole of the different mixing angles is displayed in
Fig.~\ref{fig:3+1:BR.theta14.theta34}, where we present BR($Z \to e
\mu$) and BR($Z \to \mu \tau$), respectively as a function of the 
active-sterile mixing angles, $\theta_{14}$
and $\theta_{34}$. For completeness, we single out in these plots
another observable, which is the effective neutrino mass in
neutrinoless double beta decays given in Eq.~(\ref{eq:22bbdecay}).  
Dark yellow regions correspond to values of $|m_{ee}|$ within 
future sensitivity, 
i.e. $0.01 \text{ eV} \lesssim |m_{ee}|\lesssim 0.1  \text{ eV}$ (see
Table~\ref{tab:nulesssensitivities}).  

As can be verified from 
Fig.~\ref{fig:3+1:BR.theta14.theta34}, the maximal values of  
BR($Z \to \ell_1^\mp \ell_2^\pm$) are associated with 
larger values of the
active-sterile mixing angle. (In each panel, the more dense
``diagonal'' band corresponds to contributions arising from
configurations where the active-sterile mixing angle depicted in the
$x$-axis is much larger than the other two.) 
As visible in the
left panel of Fig.~\ref{fig:3+1:BR.theta14.theta34}, for a regime of
large $\theta_{14}$, one can be indeed within reach of near
future $0\nu 2 \beta$ decay dedicated experiments (in agreement with the
findings of~\cite{Abada:2014nwa}). However, 
the associated BR($Z \to e \mu$) lies beyond 
FCC-ee expected sensitivity. Although this region would indeed be
larger in the case of an IH for the light
neutrino spectrum, the corresponding BR($Z \to e \mu$) would still remain 
below $10^{-13}$.

\begin{figure}
\begin{center}
\begin{tabular}{cc}
\epsfig{file=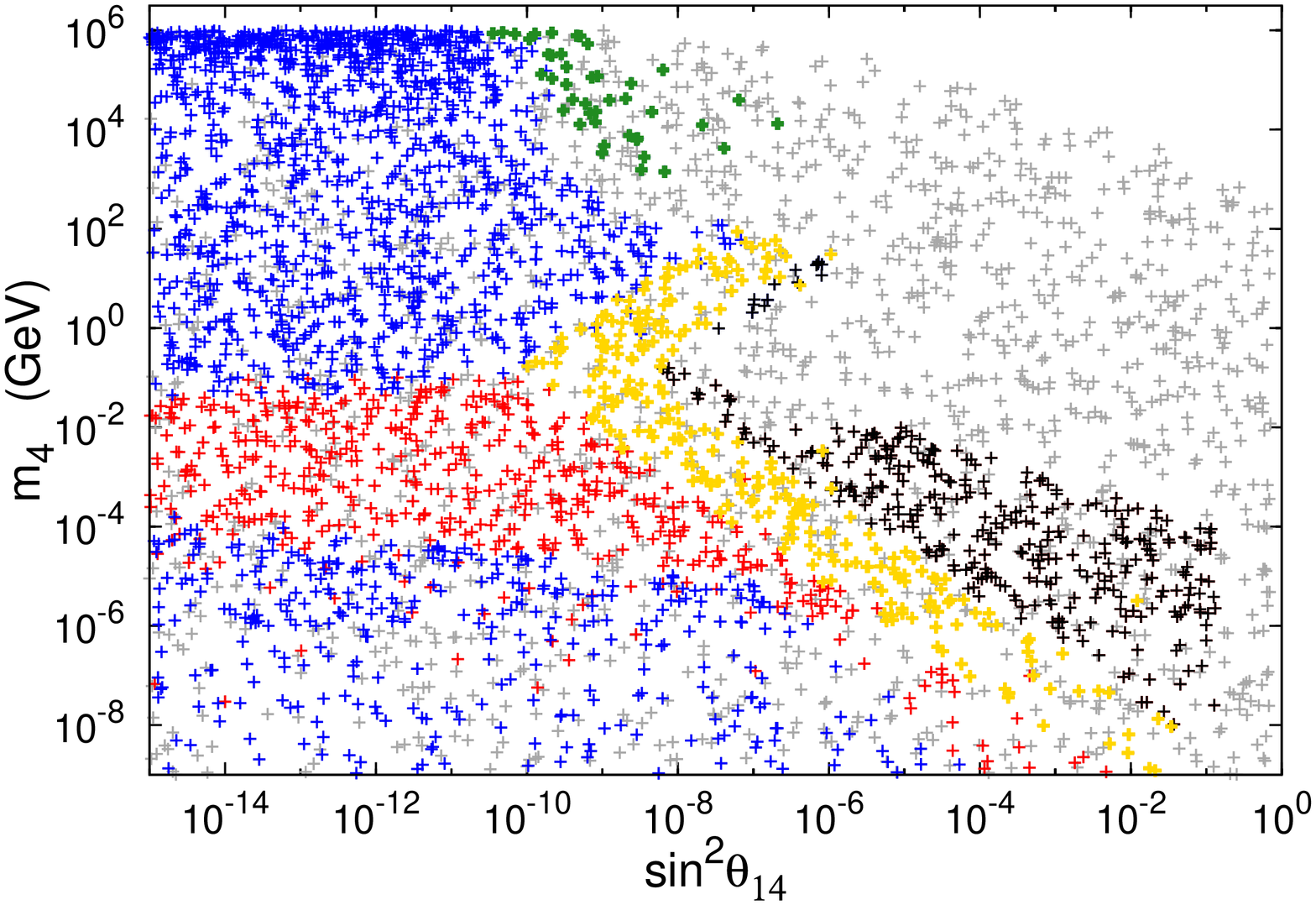, width=55mm, angle=270}
&
\epsfig{file=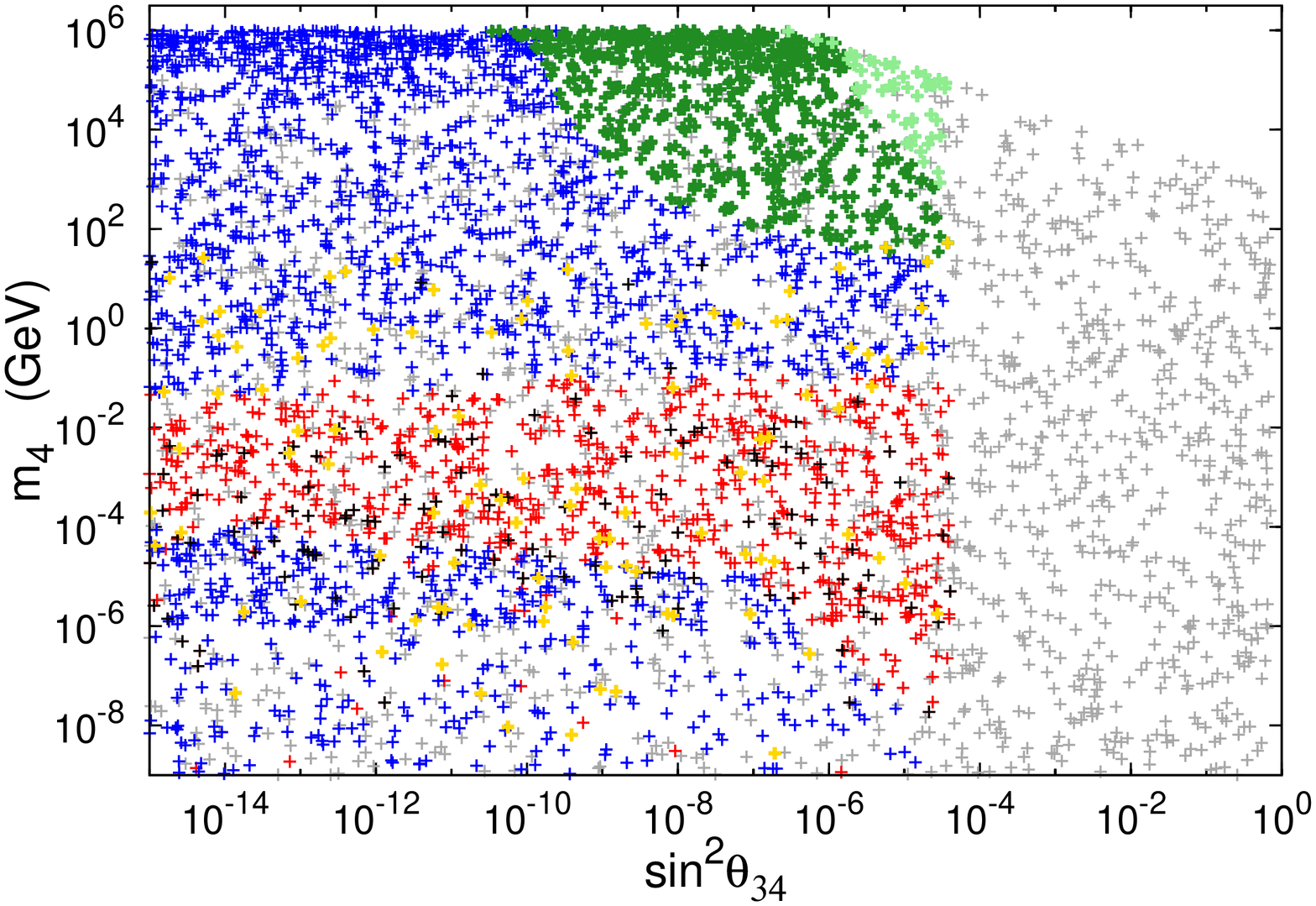, width=55mm, angle=270}
\end{tabular}
\end{center}
\caption{The ``3+1 model": on the left $(\sin^2 \theta_{14}, m_4)$ 
 parameter space of the sterile state, displaying the regimes for 
  BR($Z \to e \mu$) 
  for a NH light neutrino spectrum. Line and colour code as in
  Fig.~\ref{fig:3+1:BR.theta14.theta34} (dark green points are associated
  with $10^{-13} \lesssim $ BR($Z \to e \mu$)~$\lesssim 10^{-9} $,
  while light green ones correspond to BR($Z \to e \mu$)~$\gtrsim 10^{-9}$).
  On the right, $(\sin^2 \theta_{34}, m_4)$ displaying with the same
  colour code  
  the corresponding regimes for  BR($Z \to \mu
  \tau$). }\label{fig:3+1:m4.theta.BR} 
\end{figure}

The prospects regarding the observation of a $Z \to \ell_1^\mp
\ell_2^\pm$ decay at a
high-luminosity $Z$-factory for the full sterile neutrino parameter 
space studied in our analysis are summarised in 
Fig.~\ref{fig:3+1:m4.theta.BR}, where we display the $(\sin^2
\theta_{i4}, m_4)$ plane. (We notice that in agreement with
Eq.~(\ref{eq:sterile:bounds:Ciimi}) the upper regions, corresponding to
a regime of heavy masses and
large active-sterile mixings, are precluded due to 
  perturbativity arguments.)

As can be confirmed, and in agreement with the previous discussion,
the largest values of the lepton flavour violation $Z$-decays
correspond to regimes of large sterile masses, in association with
sizeable mixing angles. The $(\sin^2 \theta_{14}, m_4)$ parameter
space is strongly constrained by the current bounds from BR($\mu \to 3
e$) - as would be the case of $(\sin^2 \theta_{24}, m_4)$, not
displayed here - and from CR($\mu - e$, Au), while $\sin^2 \theta_{34}
\gtrsim 10^{-4}$ 
are excluded due to constraints arising from BR($\tau \to 3 \mu$).

\begin{figure}
\begin{center}
\begin{tabular}{cc}
\epsfig{file=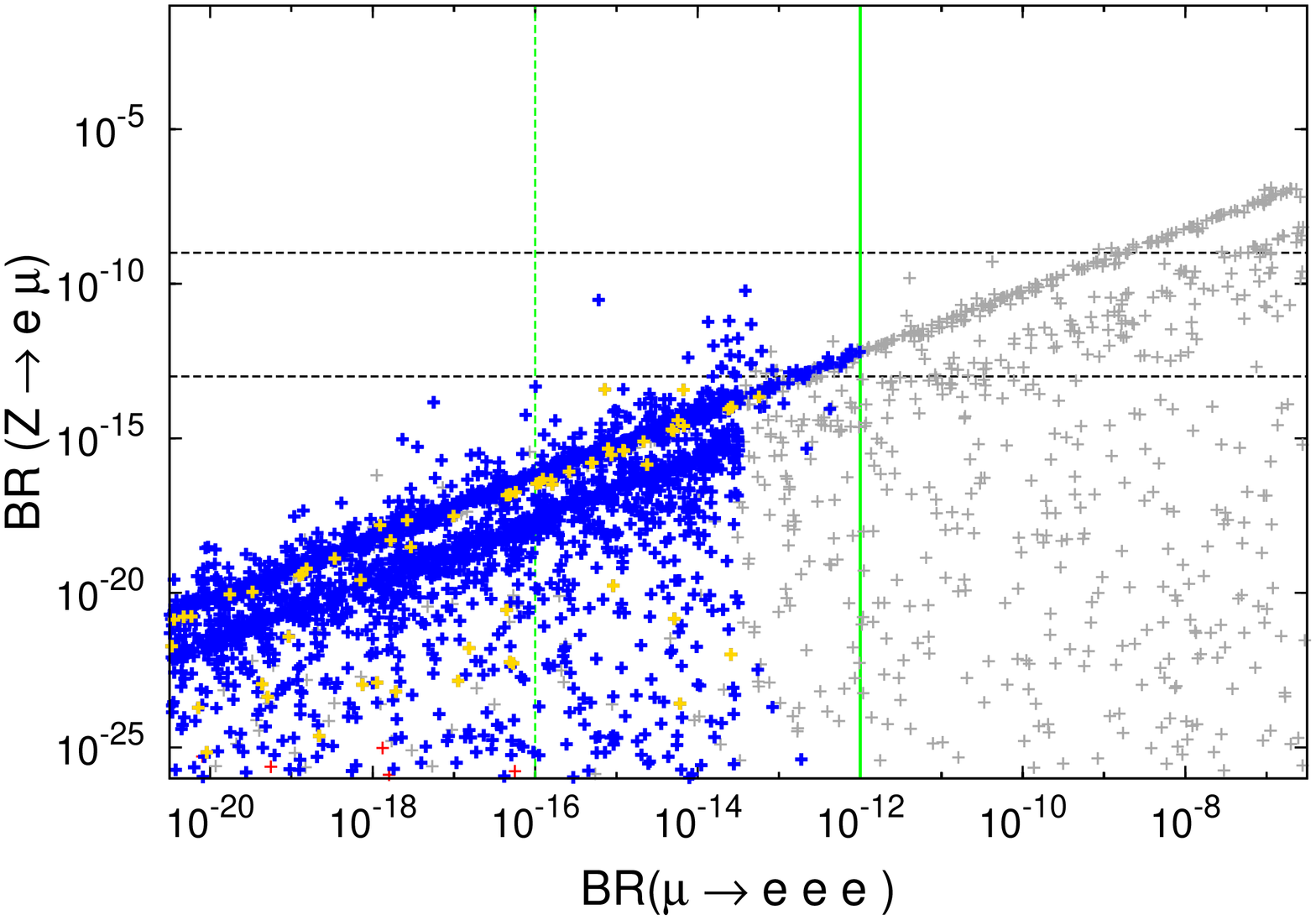, width=55mm,
  angle=270} 
&
\epsfig{file=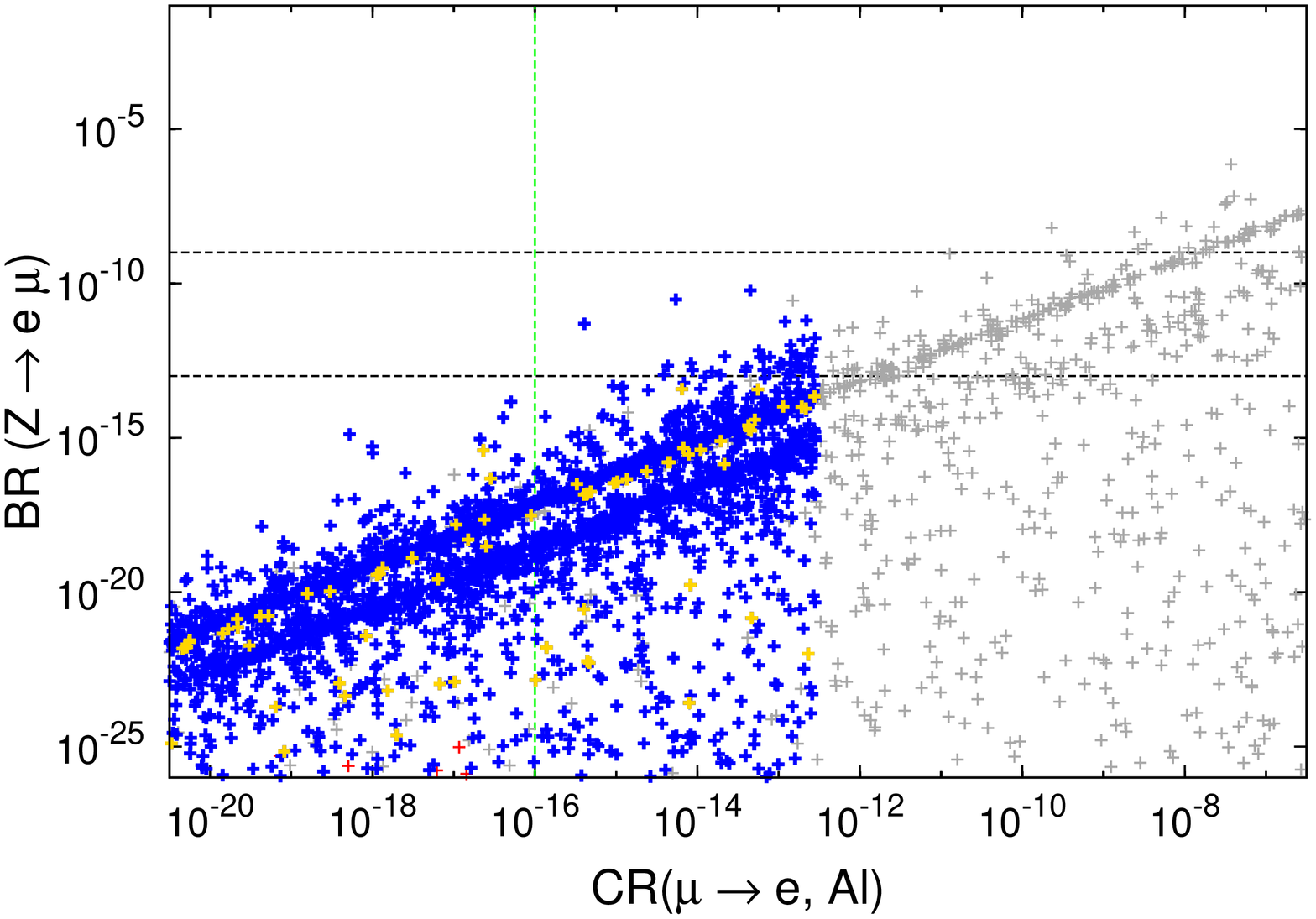, width=55mm,
  angle=270} \\ 
\epsfig{file=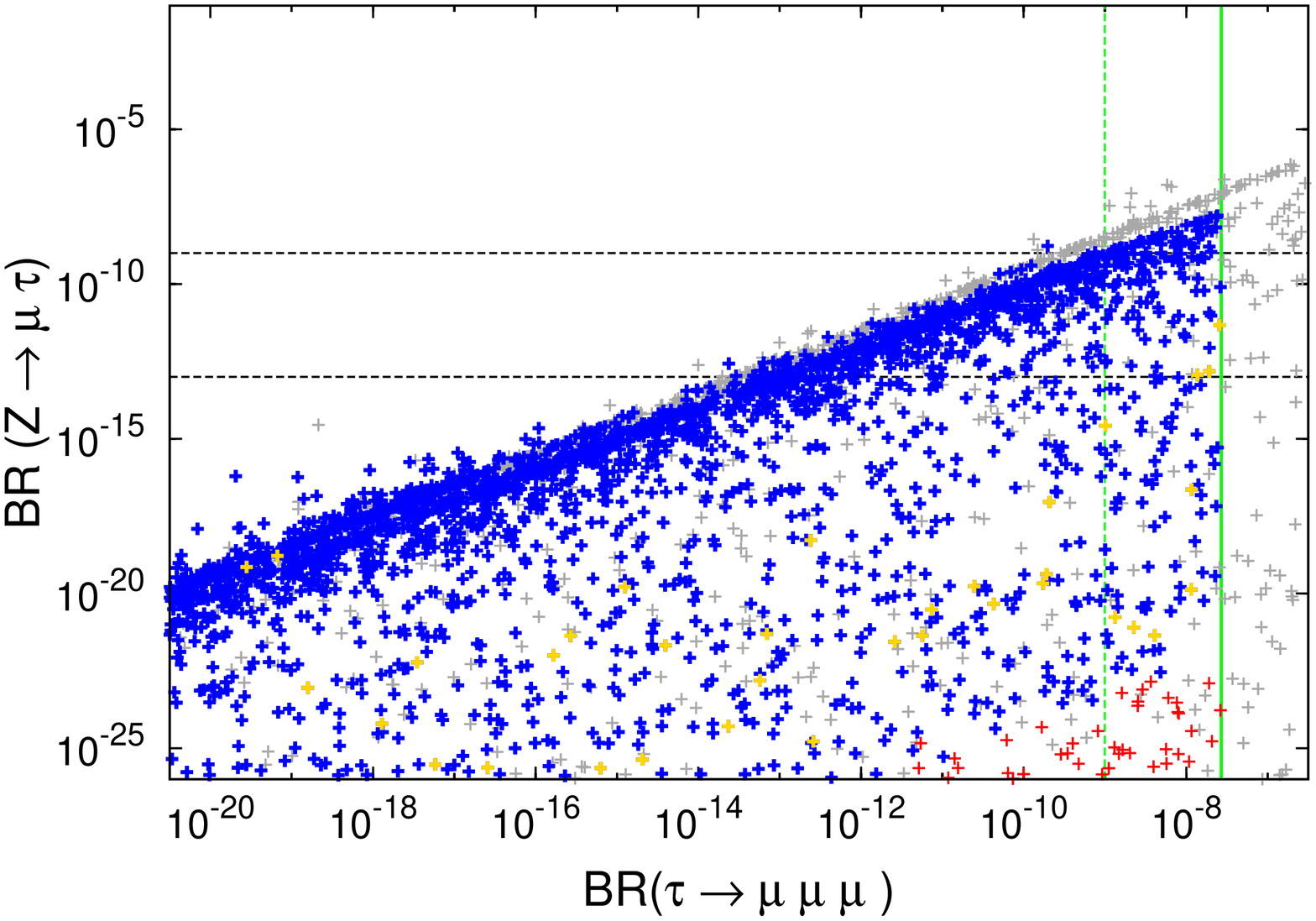, width=55mm,
  angle=270} 
&
\epsfig{file=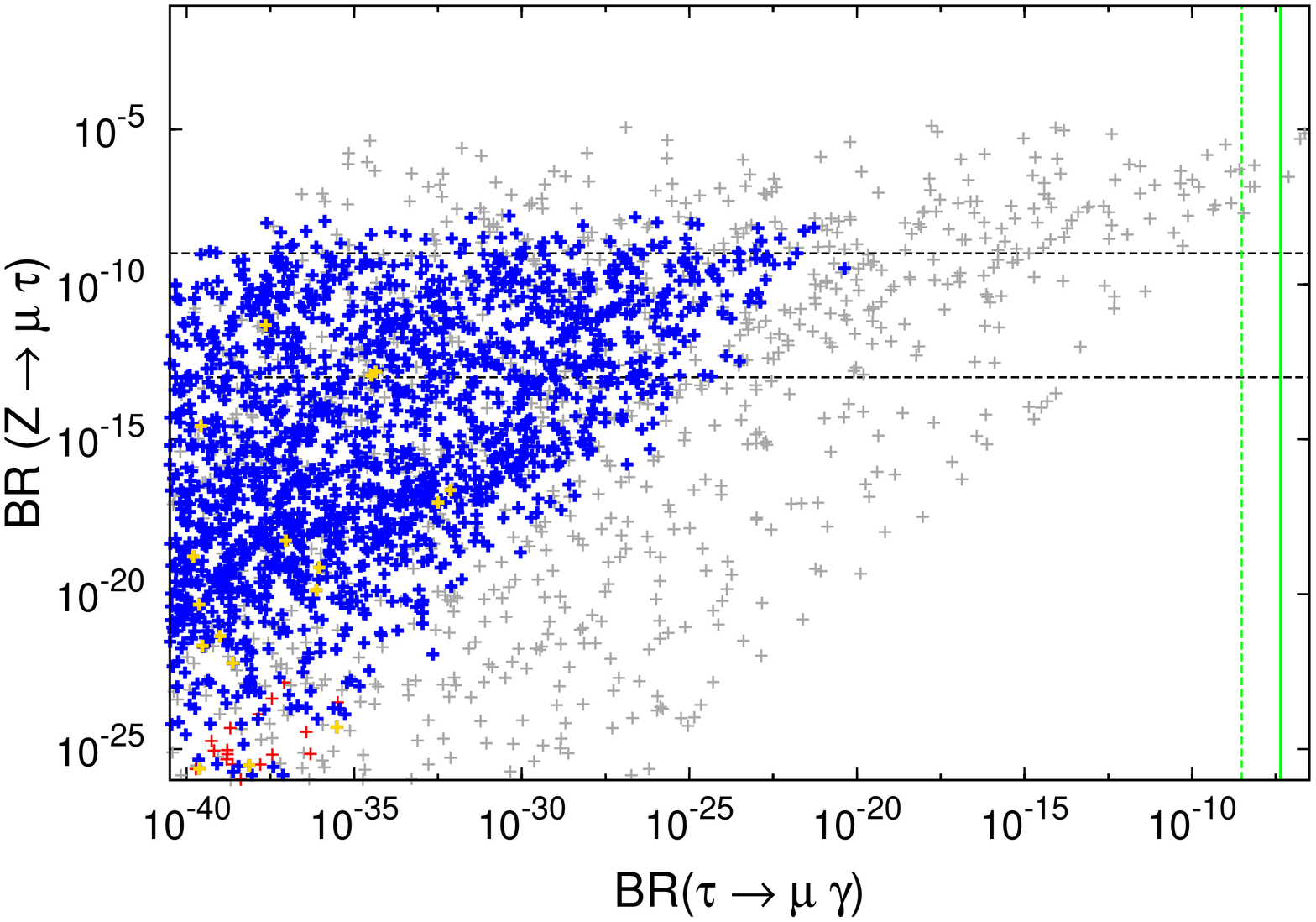,
  width=55mm, angle=270} 
\end{tabular}
\end{center}
\caption{The ``3+1 model": on the upper panels BR($Z \to e \mu$)
  versus BR($\mu \to 3 e$) (left) and CR($\mu - e$, Al) (right),
  on the lower panels BR($Z \to \mu \tau$)
  versus BR($\tau \to 3 \mu$) (left) and BR($\tau \to \mu \gamma$)
  (right) for a NH light neutrino spectrum. 
Line and colour code as in Fig.~\ref{fig:3+1:BRmt.m4}. When present, 
the additional green vertical 
lines denote the current bounds (solid) and future
sensitivity (dashed), and dark-yellow points denote an associated 
$|m_{ee}|$ within experimental reach.
}\label{fig:3+1:BRZ.low}
\end{figure}

We conclude the analysis of the ``3+1 model'' by
investigating the complementary r\^ole of a high-luminosity $Z$-factory
with respect to low-energy (high-intensity) cLFV dedicated
experiments. From the above discussion, it is clear that low-energy
cLFV processes play a constraining r\^ole in the maximal values of   
BR($Z \to \ell_1^\mp \ell_2^\pm$); we now explore which facility has
the greater potential to probe cLFV in the ``3+1 model''.
This is illustrated in Fig.~\ref{fig:3+1:BRZ.low}, where we
display the sterile neutrino contributions to BR($Z \to \ell_1^\mp
\ell_2^\pm$) versus different low-energy cLFV observables.

As can be inferred from the upper panels of
Fig.~\ref{fig:3+1:BRZ.low}, low-energy cLFV dedicated facilities offer
much better prospects to probe lepton flavour violation in the
$\mu-e$ sector of the ``3+1 model'' than a high-luminosity $Z$-factory. In
particular, Mu3e (PSI) \cite{Blondel:2013ia} and COMET (J-PARC)
\cite{Kuno:2013mha} will be sensitive to regions 
in parameter space associated with BR($Z \to e \mu$)~$\sim 10^{-17\div
  -13}$, beyond the reach of FCC-ee. Interestingly, the situation is
reversed for the case of the $\mu-\tau$ sector: as can be seen from
both lower panels of Fig.~\ref{fig:3+1:BRZ.low}, a high-luminosity
$Z$-factory such as FCC-ee allows to probe much larger regions of the 
``3+1 model'' than low-energy facilities (searching for radiative and
3-body $\tau$ decays). In particular, we draw the
attention to a small subset of the parameter space, which can be
simultaneously probed via $Z \to \mu \tau$ and $\tau \to 3 \mu$
decays, and which is also within reach of
near future $0\nu 2 \beta$  decay searches (especially in the
case of an IH light neutrino spectrum, not displayed here), opening the door
for a three-fold experimental test of this minimal SM extension.

\section{The neutrino minimal SM: $\nu$MSM}

The $\nu$MSM consists in a truly minimal extension of the SM via the
inclusion of three RH neutrinos, aiming at simultaneously
addressing the problems of neutrino mass generation, the baryon
asymmetry of the Universe (BAU) and providing a viable DM
candidate~\cite{Asaka:2005an,Asaka:2005pn,Shaposhnikov:2008pf,Canetti:2012kh}.  
In its most successful realisations, the thermally produced 
lightest sterile state accounts for the DM relic density, while the
two heavier states generate the masses of the active neutrinos. The
CP violating oscillations of the latter states produce a lepton
asymmetry via flavoured leptogenesis~\cite{Akhmedov:1998qx}, which is
converted into a baryon  
asymmetry. (For a detailed discussion, see~\cite{Asaka:2005an,Asaka:2005pn}.) 
More relaxed $\nu$MSM realisations forego a full (or partial) explanation 
of the DM relic density.

\subsection{Sterile neutrinos in the $\nu$MSM}
The addition of three generations of RH Majorana states $\nu_R$ to the
SM particle content allows to add the following terms to the leptonic
Lagrangian: 
\begin{equation}\label{eq:nuMSM:Lmass}
\mathcal{L}^\text{$\nu$MSM}_\text{mass} \, = \, 
-Y^\nu_{ij}\,\bar \nu_{Ri}\, \tilde H^\dagger L_j \, -\, 
\frac{1}{2}\, \bar \nu_{Ri}\, M_{Mij}\,\nu^c_{Rj} + \text{H.c.}\,,
\end{equation}
where $i,j=1,2,3$ are generation indices, $L$ is the SU(2)$_L$
lepton doublet and $\tilde H = i \sigma_2 H^*$; $Y^\nu$ denotes the 
Yukawa couplings, while $M_{M}$ is a Majorana mass matrix (leading to
the violation of total lepton number, $\Delta L=2$). 
After EW symmetry breaking, the neutral lepton spectrum is composed of
six Majorana fermions: the active (mostly left-handed) light states,
and three heavier sterile neutrinos.
The light neutrino masses, $m_{\nu_{1-3}}$ are given by a type I seesaw
relation\footnote{Despite the comparatively low seesaw 
  scale of the $\nu$MSM, working in 
  the seesaw limit, i.e. $m_D / M_M \ll 1$ is still a valid
  approximation.},  
\begin{equation}\label{eq:nuMSM:seesawlight}
m_{\nu_{1-3}} \, = - m_D^T \, (M_M)^{-1} \, m_D\,, \quad \text{where } \quad 
m_D \, = \, Y^\nu \,v\,, 
\end{equation}
with $v=$174 GeV the Higgs vacuum expectation value. 
The heavier spectrum, corresponding to $m_{\nu_{4-6}}$ is given by \cite{Canetti:2012kh}
\begin{equation}\label{eq:nuMSM:Nimass}
m_{\nu_{4-6}}\,=\,M_M\,+\,\frac{1}{2}\, \left(\frac{1}{M_M} (m_D^*\,m_D^T) +
(m_D^*\,m_D^T)^* \frac{1}{M_M}\right)
\end{equation}
where corrections of second order in $m_D / M_M$ are taken into account.

In order to be a good DM candidate, the couplings of $\nu_4$ to
the other active and sterile states must be very small. This
translates into associated tiny Yukawa couplings, and negligible
mixings with the heavier steriles, $\nu_5$ and $\nu_6$. 
In addition to light neutrino mass generation
(in which $\nu_4$ plays no r\^ole), the latter two
states, are responsible for generating
lepton asymmetries: on the one hand, the asymmetries produced at
early times will give rise to BAU, while those at late times can account
for the correct rate of thermal $\nu_4$ production~\cite{Laine:2008pg}. 
In both cases, the leptonic asymmetry generation in general 
relies on a resonant
amplification \cite{Pilaftsis:2003gt}, and the heavier steriles,
$\nu_5$ and $\nu_6$, exhibit a 
certain amount of degeneracy. 

There are several possible parametrizations of the physical $\nu$MSM
degrees of freedom. 
Drawing from the analysis of the ``3+1 model'' discussed in
Section~\ref{sec:3+1}, we prefer to carry our discussion in
terms of the six mass eigenvalues, while encoding all physical 
mixing angles and CP violating phases (Dirac and Majorana) in an 
effective $6\times 6$ unitary mixing matrix, ${\bf U}$, 
as it allows to readily implement
the already well-established bounds on the $\nu$MSM parameter space. 
The angles $\theta_{lj}$, $l=1,2,3$, $j=4,5,6$ encode the
active-sterile mixings, while the mixings between the sterile states
are given by three additional angles $\theta_{45,46,56}$. The matrix
${\bf U}$ is further parametrized by 3 additional Majorana and 9
Dirac phases.  The heavier masses can be written as:
\begin{equation}\label{eq:nuMSM:massN}
m_{j} \, = \, \operatorname{diag}(m_\text{DM}, M - \delta_M, M + \delta_M) \,
\end{equation}
with $j=4,5,6$. In the above $m_4 = m_\text{DM}$ is the mass of the
DM candidate.

\bigskip
In addition to the general constraints on sterile neutrino extensions
of the SM, the peculiar features of the $\nu$MSM (generation of the
BAU and a viable DM candidate) lead to a very constrained parameter
space. Here we rely on the results
of~\cite{Canetti:2012kh}, where the most 
relevant constraints are translated into bounds on the $(U^2, M)$
planes, as well as on the splitting $\delta_M$, which is of the order $\sim 10^{-4}~\rm eV - 1~\rm keV$.
The quantity $U^2$ encodes the experimentally relevant
combination of couplings; in the limit of small active-sterile
mixings, and in analogy to~\cite{Canetti:2012kh}, we will use
\begin{align}\label{eq:nuMSM:Utheta}
& U^2_{4}\,=\, U^2_{e4} + U^2_{\mu4} + U^2_{\tau4} \,=\,\sum_{l} \sin^2 \theta_{l4}, \nonumber \\
& U^2_{4-6}\,=\,\sum_{l}  \sin^2 \theta_{l4} + \sin^2 \theta_{l5}
+ \sin^2 \theta_{l6}\,, \quad \text{with}\, l = e,\mu, \tau. 
\end{align}

\noindent
{\bf Dark matter constraints}

\noindent
As reported in~\cite{Canetti:2012kh}, observations of the matter
distribution in the 
Universe constrain the DM free streaming length; realistic scenarios
(including combinations of X-ray bounds and Ly$_\alpha$ forest
reconstruction, among others) suggest 
$10 \text{ keV} \lesssim m_4  \lesssim 50 \text{ keV} $;
combining the latter bounds with a successful 
production of the required DM abundance, one is led to bounds on the
corresponding mixing angles, $\theta^2_{l4} \sim \mathcal{O} (10^{-13}
- 10^{-8})$, $l=1,2,3$. 
In our analysis we will not exclude regions in which the lightest
sterile would have a relic density below the observed value (i.e.,
smaller values of $\theta^2_{l4}$).
\bigskip

\noindent
{\bf Heavy sterile parameter space}

\noindent
As discussed in~\cite{Canetti:2012kh}, the allowed  
$(U^2_{4-6}, M)$ plane corresponds to a well-defined region: 
the regime of very small mixings is excluded by the impossibility of 
correctly reproducing the active neutrino mass differences (seesaw
exclusion), while
larger mixings preclude the generation of a baryon asymmetry from
RH neutrino oscillations; the BAU exclusion surface extends
to the seesaw exclusion, effectively constraining the average
$\nu_{5,6}$ masses to lie below the EW scale. Finally, the small mass
regime (i.e., $m_{\nu_{5,6}} \lesssim 0.1$ GeV) is also ruled out due to
conflict with BBN bounds and direct searches (at
PS191~\cite{Bernardi:1985ny,Bernardi:1987ek}).  
Although we have used the BAU-derived constraints on the magnitude of
the distinct mixing angles, we have not attempted at doing the same
for the CP violating phases, which for simplicity were set to zero in
this analysis. 

\begin{figure}
\begin{tabular}{cc}
\epsfig{file=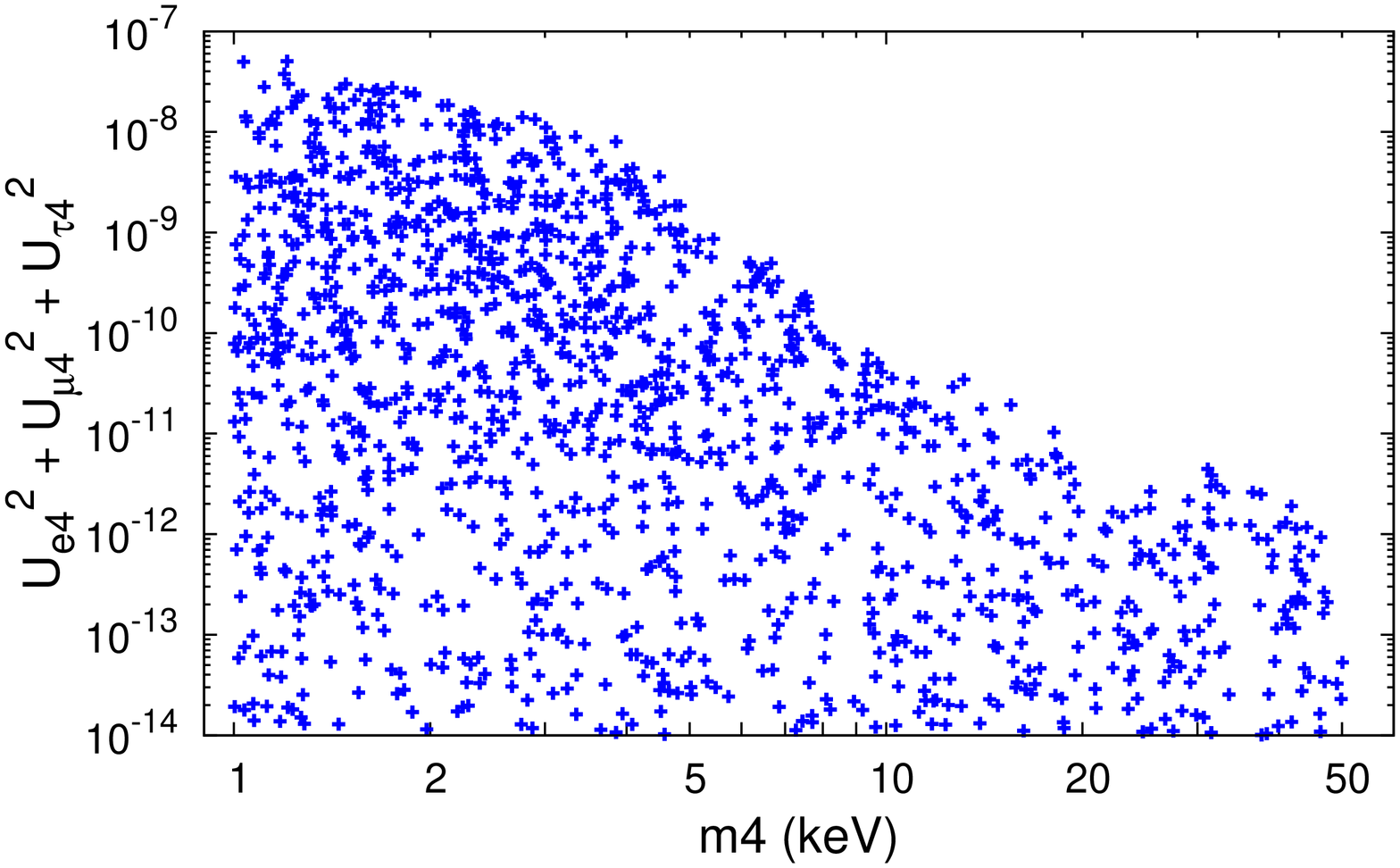, width=55mm, angle=270}
&
\epsfig{file=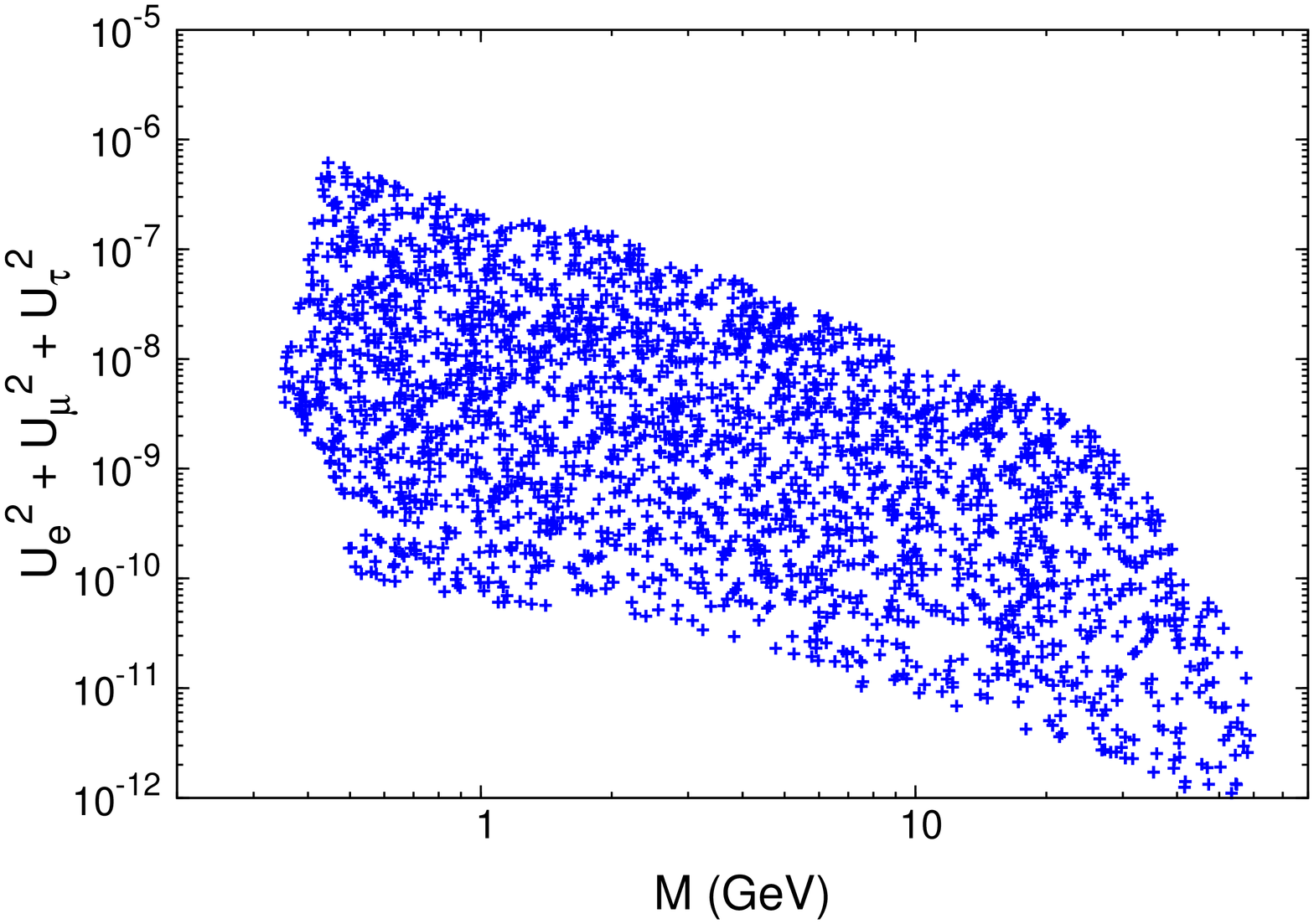, width=55mm, angle=270}
\end{tabular}
\caption{$\nu$MSM model: $(U^2_{4} , m_\text{DM})$ 
and $(U^2_{4-6} ,  M)$ parameter spaces (as identified
in~\cite{Canetti:2012kh}), 
respectively on the left and right panels, for a NH light neutrino spectrum. 
}\label{fig:nuMSM:N1.N23}
\end{figure}

In Fig.~\ref{fig:nuMSM:N1.N23}, and for completeness, 
we summarise the $\nu$MSM
parameter space investigated in our analysis, closely following the
dedicated studies of~\cite{Canetti:2012kh}, and assuming a NH for the
light neutrino spectrum. 

\subsection{Leptonic $Z$ decays in the $\nu$MSM}
\begin{figure}
\begin{tabular}{cc}
\epsfig{file=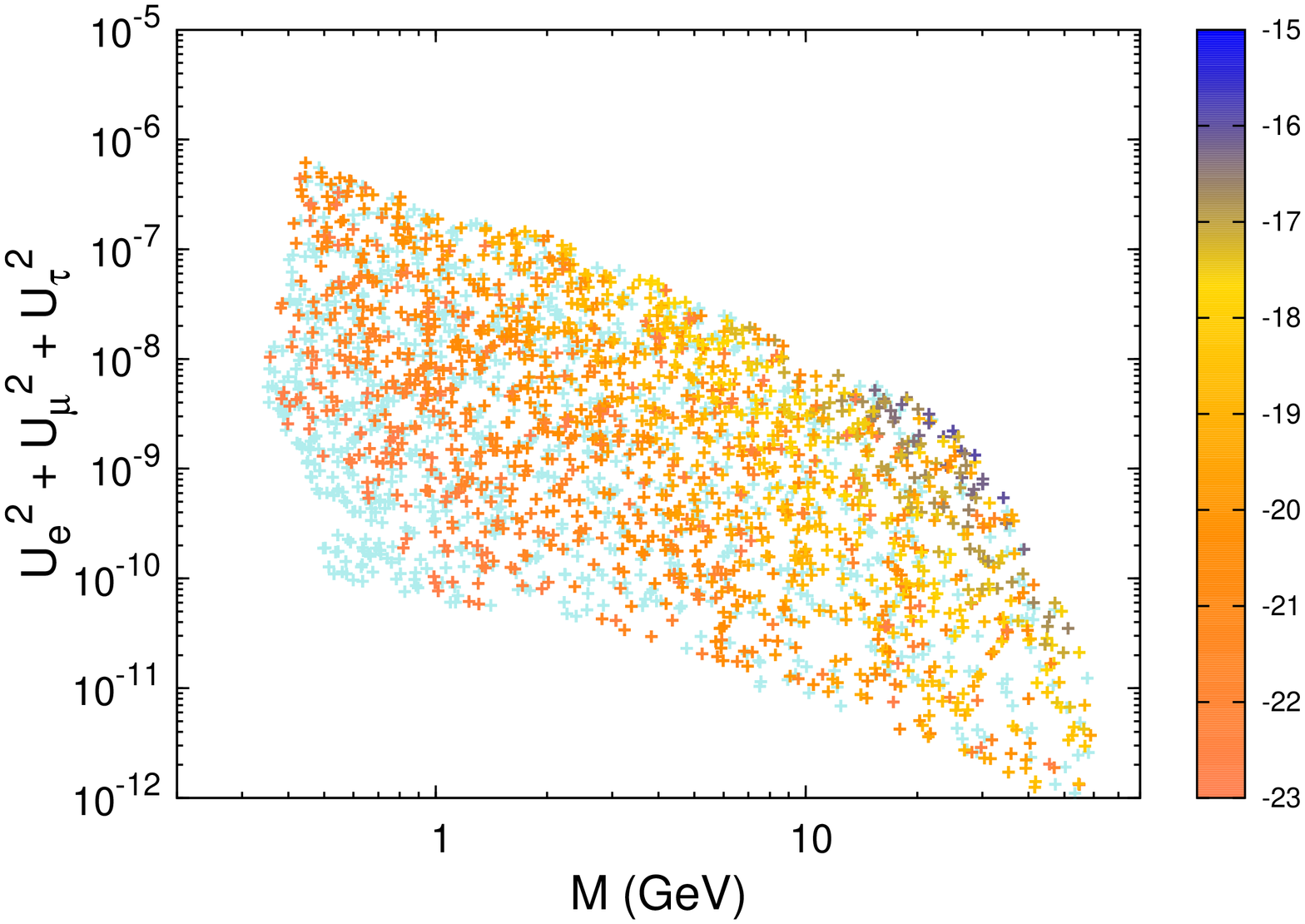, width=55mm, angle=270}
&
\epsfig{file=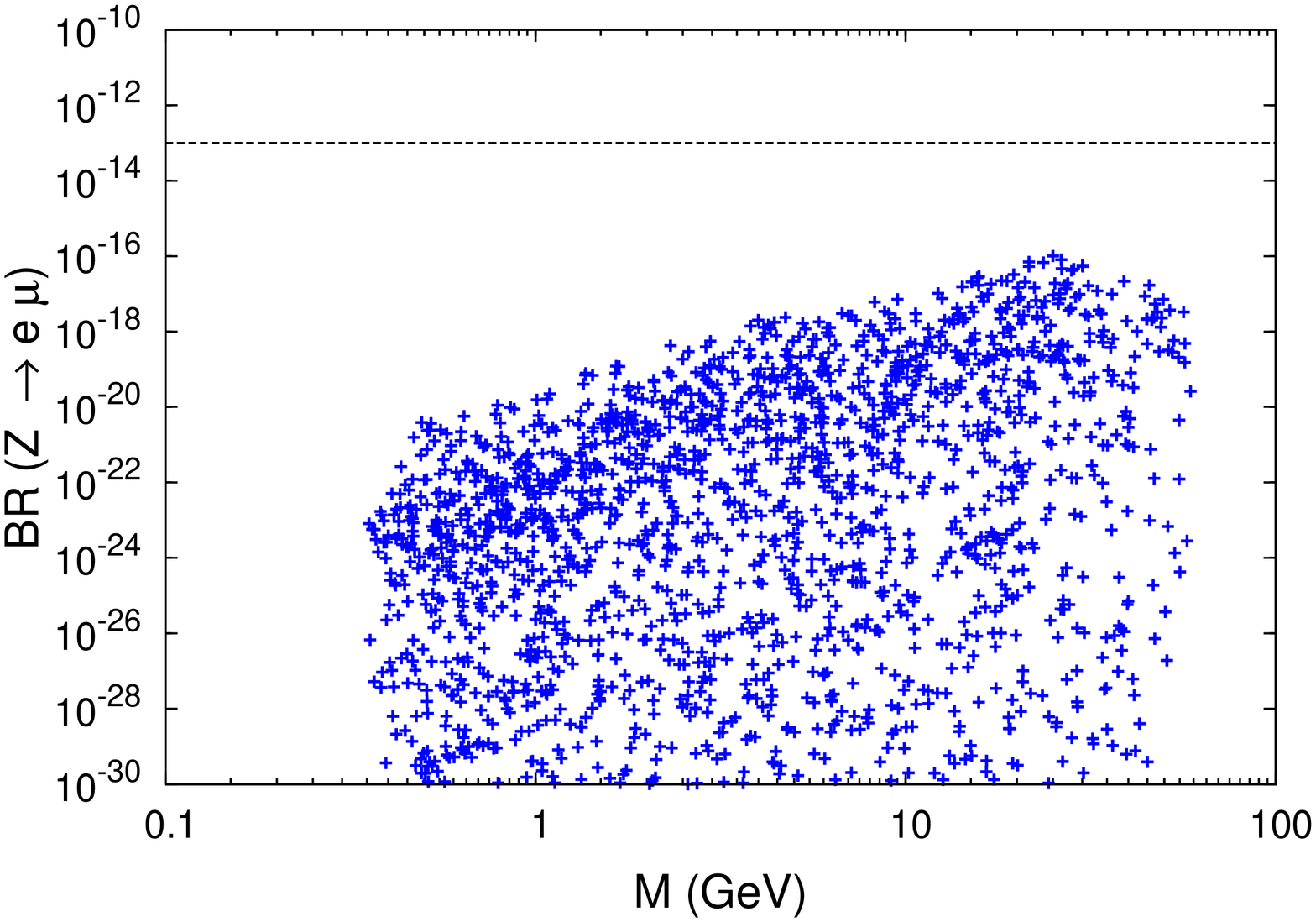, width=55mm, angle=270}
\end{tabular}
\caption{$\nu$MSM model: on the left, maximal values of 
  BR($Z \to \ell_1^\mp \ell_2^\pm$) on the 
  $(U^2_{4-6},M)$ parameter space, from larger (dark blue) to smaller
  (orange) values. Cyan denotes values of the branching fraction below
  $10^{-23}$. On the right BR($Z \to e \mu$) as a
  function of $M$, for the points in the allowed $(U^2_{4-6},
  M)$ parameter space. Both cases correspond to a NH light neutrino spectrum. 
}\label{fig:nuMSM:U.M.BR:BR.M}
\end{figure}

We begin by discussing the expected 
BR($Z \to \ell_1^\mp \ell_2^\pm$) within the $\nu$MSM. 
In Fig.~\ref{fig:nuMSM:U.M.BR:BR.M}, we
display the range of LFV $Z$ boson decays across the allowed parameter space.

As expected from the results of Section~\ref{sec:3+1}, the maximal
values of BR($Z \to \ell_1^\mp \ell_2^\pm$) occur for a regime where sizable
RH neutrino masses are accompanied by the maximally allowed 
active-sterile mixings. Nevertheless, and as can be directly inferred
from the left panel of Fig.~\ref{fig:nuMSM:U.M.BR:BR.M}, one can have,
at best, BR($Z \to \ell_1^\mp \ell_2^\pm$)~$\lesssim
\mathcal{O}(10^{-16})$. Larger values would indeed be possible, but 
are excluded by the requirement of
generating the observed BAU. A clearer insight can be drawn from
the right panel of Fig.~\ref{fig:nuMSM:U.M.BR:BR.M}, where one
verifies that, for instance, BR($Z \to e \mu$)~$\lesssim
\mathcal{O}(10^{-16})$. Similar ranges are obtained for the other LFV
final states. Although we do not display the corresponding analysis
here, we have numerically verified that similar results are obtained
for a IH light neutrino spectrum. 
We also notice that the ranges for the LFV $Z$-decays BRs are in fair
agreement\footnote{
It is worth mentioning that our study of the $\nu$MSM - based on a
``3+3'' analysis  along the lines of the ``3+1 toy model''- leads to a
conservative estimate of the corresponding BR($Z \to \ell_1^\mp \ell_2^\pm$). 
The effective $6 \times 6$ unitary mixing matrix whose entries are
thus scanned allows to cover, and even go beyond, regions of parameter
space strictly arising in the type-I seesaw of the $\nu$MSM. Hence, we 
are not under-estimating the cLFV $Z$ decays.}
with the analysis carried for the truly minimal ``3+1
model'' in Section~\ref{sec:3+1}, considering the appropriate mass
and sterile mixing regime. 

Regarding the departure from unitarity of the $\tilde U_\text{PMNS}$ matrix
in the surveyed parameter space, we notice that (as expected) it is
comparatively small: $\tilde \eta \lesssim 10^{-6}$. 
In what concerns low-energy (charged) lepton flavour observables, due
to the smallness of the active-sterile mixings, the contributions are
typically very small, as already suggested in~\cite{Canetti:2013qna} 
regarding $\mu-e$ conversion in Nuclei.
Finally, and concerning the violation of lepton universality in
$Z$-decays, the contributions of the new sterile states of the 
$\nu$MSM are truly negligible.

For completeness, we summarise in Table~\ref{table:numsm:points} two
examples of points in the $\nu$MSM parameter space that would account
for ``maximal'' values of BR($Z \to \ell_1^\mp \ell_2^\pm$).

\begin{center}
\hspace*{5mm}
\begin{table}{\small
\begin{tabular}{|c|c|c|c|c|c|c|}
\hline
$m_4$ (keV) & $M$ (GeV) & $U^2_{4}$ & $U^2_{4-6}$ & 
BR($Z \to e \mu$) & BR($Z \to \mu \tau $) &BR($Z \to e \tau$)\\
\hline
11.8 & 26.2 & 4 $\times 10^{-25}$ & 1.8 $\times 10^{-9}$ & 
$10^{-16}$ &  7 $\times 10^{-18}$ & 2 $\times 10^{-21}$ \\
1.1 & 34.4 & 1.3 $\times 10^{-16}$ & 5.4 $\times 10^{-10}$ &
3 $\times 10^{-25}$ &  2 $\times 10^{-17}$ & 8 $\times 10^{-17}$ \\
\hline
\end{tabular}\caption{Example of two points in $\nu$MSM parameter space with
  associated BR($Z \to \ell_1^\mp \ell_2^\pm$)~$\gtrsim
  10^{-17}$.}\label{table:numsm:points}} 
\end{table}
\end{center}

It has been recently pointed out that high-luminosity $Z$-factories
(such as FCC-ee) offer a promising set-up for direct searches of RH
(nearly) sterile neutrinos, as those present in the framework of the
$\nu$MSM~\cite{Blondel:2014bra}. The small active-sterile mixing
angles lead to long lifetimes, with decay lengths comprised between 100
microns and 5m; this would allow to cover a large region
of the phase-space for heavy neutrino masses between 10 and 80 GeV,
reaching down to a mixing as small as $U^2_{4-6} \approx 10^{-12}$ (thus
complementing~\cite{Blondel:2014bra} the probing power of the SHIP
experiment~\cite{Bonivento:2013jag}).  
Lepton flavour violating $Z$ decays do not allow a further synergy with
the above mentioned searches for light RH neutrinos, as those present
in the $\nu$MSM. However, the observation of LFV $Z$ decays at a 
high-luminosity $Z$-factory would suggest that sources of LFV
- other than the $\nu$MSM - are present. Conversely, the
interpretation of displaced vertices in association of a long-lived RH
state of the $\nu$MSM should not be accompanied by a  
BR($Z \to \ell_1^\mp \ell_2^\pm$) within FCC-ee sensitivity.

\section{The Inverse Seesaw scenario}
The Inverse Seesaw mechanism~\cite{Mohapatra:1986bd} 
consists in an appealing extension of the
SM via RH and sterile neutrinos. Contrary to most (type I)
low-energy seesaw realisations, the ISS allows to accommodate
neutrino data with natural values of the Yukawa couplings for a
comparatively low seesaw scale. The possibility of having sizeable 
mixings between the active and sterile states renders the model
phenomenologically rich, with a potential impact for a number of
observables.  

Depending on its actual realisation, the ISS does allow to accommodate
the observed DM relic abundance and (potential) indirect DM
detection hints~\cite{Abada:2014vea,Abada:2014zra}.

\subsection{The (3,3) ISS realisation}
In the ISS, $n_R \ge 2$ generations of RH neutrinos 
$\nu_R$ and $n_X$ generations of extra $SU(2)$ singlets fermions $X$ (such that 
$n_R+n_X = n_s$), are added to the SM content.  Both $\nu_R$ and $X$
carry lepton number $L=+1$~\cite{Mohapatra:1986bd}.
Here we consider a specific ISS realisation in which $n_R = n_X = 3$,
the so-called (3,3) realisation. The SM Lagrangian is thus extended as
\begin{equation}
\mathcal{L}_\text{ISS} \,=\, 
\mathcal{L}_\text{SM} - Y^{\nu}_{ij}\, \bar{\nu}_{R i} \,\tilde{H}^\dagger  \,L_j 
- {M_R}_{ij} \, \bar{\nu}_{R i}\, X_j - 
\frac{1}{2} {\mu_X}_{ij} \,\bar{X}^c_i \,X_j + \, \text{h.c.}\,,
\end{equation}
where $i,j = 1,2,3$ are generation indices and $\tilde{H} = i \sigma_2
H^*$. Notice that $U(1)_L$ (i.e., lepton number) is broken
only by the non-zero Majorana mass term $\mu_{X}$, while the Dirac-type 
RH neutrino mass term $M_{R}$ does conserve lepton number.
In the $(\nu_L,{\nu^c_R},X)^T$ basis, and after EW symmetry breaking, 
the (symmetric) $9\times9$ neutrino mass matrix 
$\mathcal{M}$ is given by
\begin{eqnarray}
{\cal M}&=&\left(
\begin{array}{ccc}
0 & m^{T}_D & 0 \\
m_D & 0 & M_R \\
0 & M^{T}_R & \mu_X \\
\end{array}\right) \, ,
\label{eq:ISS:M9}
\end{eqnarray}
with $m_D= Y^\nu v$ the Dirac mass term,  
$v$ being the vacuum expectation value of the SM Higgs boson.  
Under the assumption that $\mu_X \ll m_D \ll M_R$, the
diagonalization of ${\cal M}$ leads to an effective Majorana mass
matrix for the active (light) neutrinos~\cite{GonzalezGarcia:1988rw},
\begin{equation}\label{eq:nu}
m_\nu \,\simeq \,{m_D^T\, M_R^{T}}^{-1} \,\mu_X \,M_R^{-1}\, m_D \, .
\end{equation}
The remaining (mostly) sterile states form nearly degenerate pseudo-Dirac
pairs, with masses 
\begin{equation}\label{eq:ISS:pseudodirac}
 m_{S_\pm}=\pm \sqrt{M_{R}^2+m_{D}^2} + \frac{M_{R}^2 \,\mu_X}{2\,
   (m_{D}^2+M_{R}^2)}\,.  
\end{equation}
It proves convenient to introduce the following matrix
$M = M_R\, \mu_X^{-1} \,M_R^T$, which is diagonalized 
as $D M D^T = \hat{M}$. The eigenvalues of $M$ are thus the
entries of the diagonal matrix $\hat M$. 
In order to write the neutrino Yukawa couplings, it is useful to 
use a generalization of the Casas-Ibarra parametrization~\cite{Casas:2001sr}, 
which allows to cast $Y^\nu$ as

\begin{equation}\label{eq:YvcasasI}
Y^\nu \,= \,
\frac{1}{v} \, D^\dagger \, \sqrt{\hat M} \, R \,
\sqrt{{\hat m}_\nu} \, U_\text{PMNS}^\dagger \, . 
\end{equation}

In the above, $\sqrt{{\hat m}_\nu}$ is a diagonal matrix containing the square
roots of the three light neutrino mass eigenvalues $m_\nu$, $R$ is an
arbitrary $3 \times 3$ complex 
orthogonal matrix, parametrized by $3$ complex angles, encoding the
remaining degrees of freedom. (Without loss of generality, one can
choose to work
in a basis where $M_R$ is a real diagonal matrix, as are the
charged lepton Yukawa couplings.)
The full neutrino mass matrix is then diagonalized by the $9\times 9$
unitary mixing matrix ${\bf U}$ as ${\bf U}^T \mathcal{M} {\bf U} 
= \text{diag}(m_i)$. In the basis where the charged lepton mass matrix is
diagonal, the leptonic mixing matrix is given by the
rectangular $3 \times 9$ sub-matrix corresponding to the first three
columns of ${\bf U}$, with the $3 \times 3$ block corresponding to the 
(non-unitary\footnote{For further studies on
non-unitarity effects in the Inverse Seesaw see, for instance, \cite{Forero:2011pc,Malinsky:2009gw,Dev:2009aw}.}) $\tilde U_\text{PMNS}$.

\bigskip
In the following numerical study, the contributions to the distinct
observables are derived 
through the following general scan: leading to the construction of the
$9\times 9$ mass matrix in Eq.~(\ref{eq:ISS:M9}),
the modulus of the entries of the matrices $M_R$ and $\mu_X$ are
randomly taken to lie on the intervals
$0.1 \text{ MeV} \lesssim (M_R)_{i}  \lesssim 10^6 \text{ GeV}$ and
$0.01 \text{ eV} \lesssim
(\mu_X)_{ij}  \lesssim 1 \text{ MeV}$, with complex entries for the 
lepton number violating matrix $\mu_X$; 
we also take complex angles for the $R$ matrix, randomly 
varying their values in the interval $[0, 2\pi]$.
The modified Casas-Ibarra parametrization for $Y^\nu$,
Eq.~(\ref{eq:YvcasasI}), ensures that constraints from 
neutrino oscillation data are 
satisfied. 

\subsection{ISS: Violation of flavour universality in $Z$ decays}
Despite the contributions of the several additional states of the ISS to the
violation of lepton flavour universality observable $\Delta
R^\text{lep}_Z$, see Eq.~(\ref{eq:DeltaZ}), the ISS also
remains short of the future sensitivity. 
Although in regions of the surveyed parameter space one
could in principle have  $\Delta R^\text{lep}_Z \sim 10^{-3}$, this
regions are experimentally excluded, as there are strong conflicts
with numerous bounds, especially those arising from low-energy 
cLFV observables.

\subsection{LFV $Z$ decays in the ISS}
\begin{figure}
\begin{tabular}{cc}
\epsfig{file=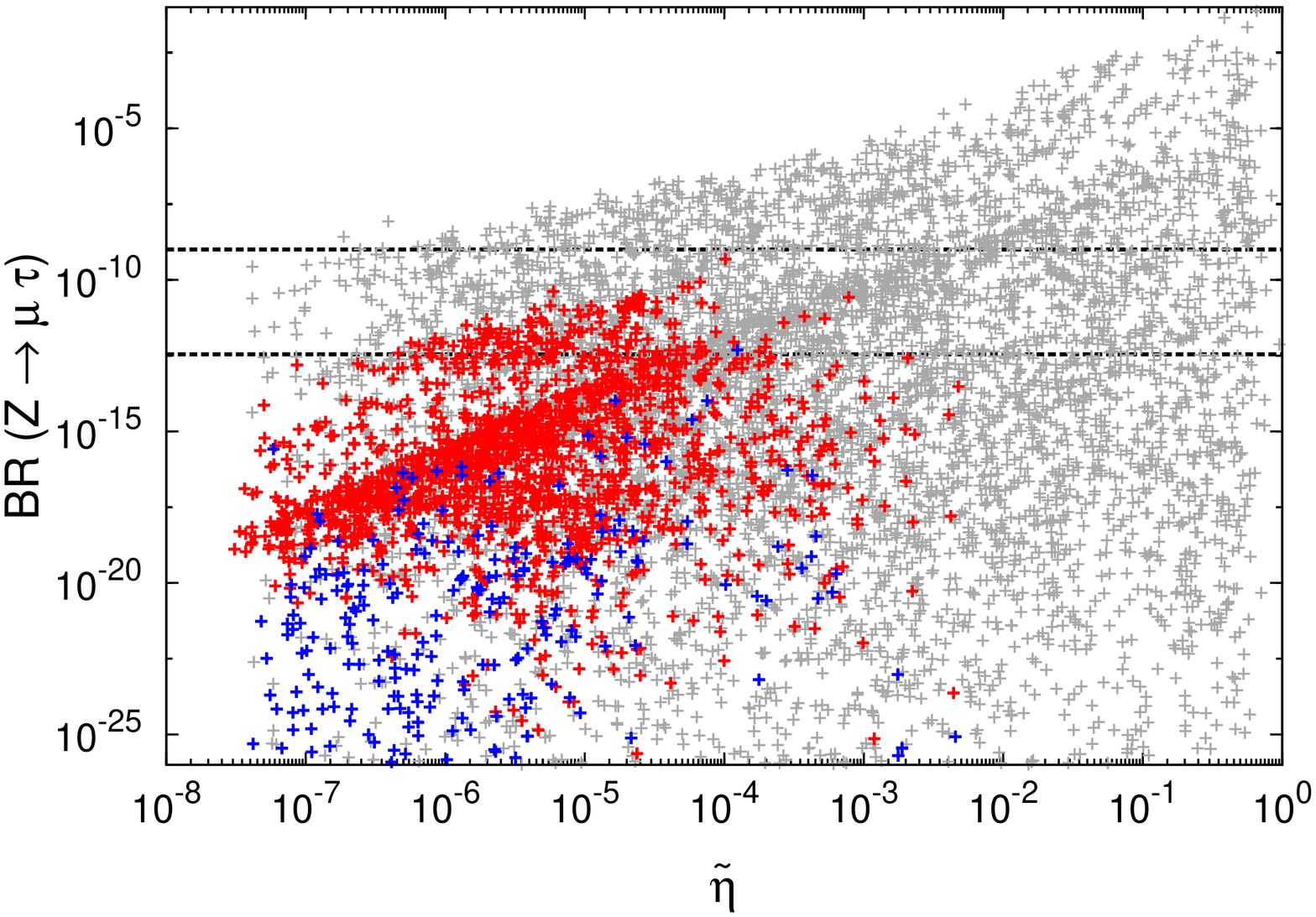, width=55mm, angle=270}
&
\epsfig{file=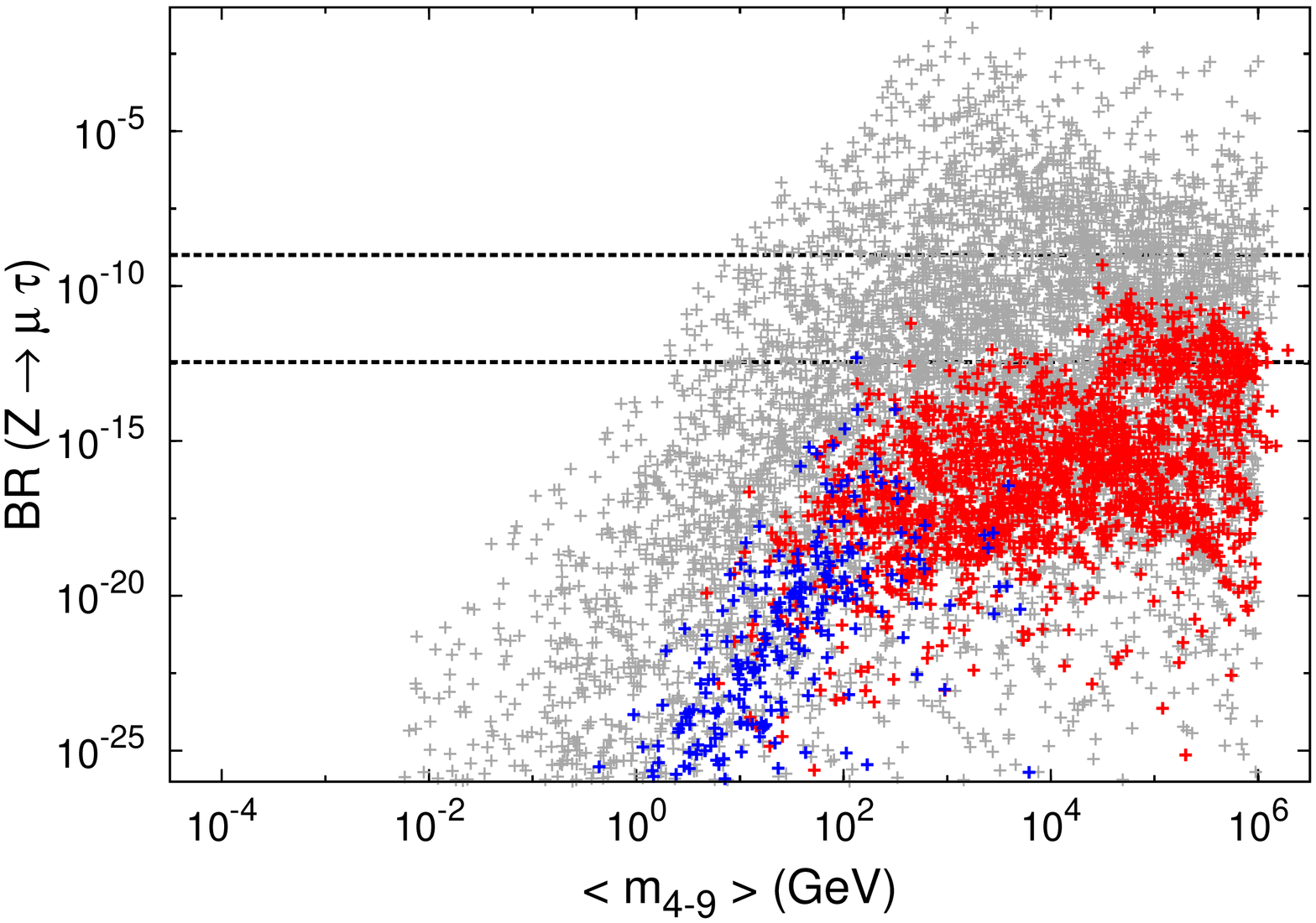, width=55mm, angle=270}
\end{tabular}
\caption{ISS realisation: BR($Z \to \mu \tau$) as a function of
 $\tilde \eta$ (left) and of the average value of the mostly
  sterile state masses (right), $\langle m_{4-9} \rangle$, 
  for a NH light neutrino
  spectrum. Line and colour code as in Fig.~\ref{fig:3+1:BRmt.m4}.
}\label{fig:ISS:BRmt.eta.avmass}
\end{figure}

We first consider the LFV decays 
$Z\to \mu \tau$, displaying the corresponding BRs on 
Fig.~\ref{fig:ISS:BRmt.eta.avmass}
as a function of $\tilde \eta$ (see Eq.~(\ref{eq:def:etatilde})), and
as a function of the average of
the absolute masses of the mostly sterile states, 
\begin{equation}\label{eq:ISS:mee:avmass}
\langle m_{4-9} \rangle\,=\, \sum_{i=4...9} \frac{1}{6}\,|m_i|\,.
\end{equation}
The results collected in Fig.~\ref{fig:ISS:BRmt.eta.avmass} reveal that
the present ISS realisation can account for sizeable values of LFV
$Z$-decay branching ratios: this in general requires the presence of
sterile states with a mass~$\gtrsim \Lambda_\text{EW}$, and can occur
even for very mild deviations from unitarity of the $\tilde
U_\text{PMNS}$. Other LFV decays, $Z\to e \mu$ and  $Z\to e \tau$ 
have somewhat smaller BRs $\lesssim \mathcal{O}(10^{-11})$, 
but still within experimental sensitivity.
(Notice that points with associated BR($Z \to \ell_1^\mp
\ell_2^\pm$) within FCC-ee reach are cosmologically disfavoured in contrast to what was encountered in the study of the simple toy model of Section \ref{sec:3+1}.)
Again, even though we only display
the NH for the light neutrino spectrum, our numerical results show
that the corresponding prospects for BR($Z \to \ell_1^\mp \ell_2^\pm$) 
would be similar in an IH case.
 
\begin{figure}
\begin{center}
\epsfig{file=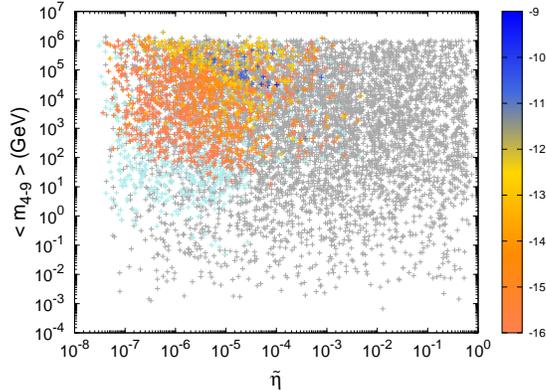, width=55mm,
  angle=270} 
\end{center}
\caption{ISS realisation: maximal values of 
  BR($Z  \to \ell_1^\mp \ell_2^\pm$) 
  on the $(\tilde \eta,\langle m_{4-9} \rangle)$ 
  parameter space for a NH light neutrino spectrum, from larger 
  (dark blue) to smaller (orange) values. 
  Cyan denotes values of the branching fractions below
  $10^{-18}$. 
}\label{fig:ISS:eta.ave.BR}
\end{figure}

Just as previously done, we summarise the prospects for the
observation of cLFV $Z$ decays in the framework of the ISS by
considering the $(\tilde \eta, \langle m_{4-9} \rangle)$ 
parameter space of this specific realisation; this is illustrated in
Fig.~\ref{fig:ISS:eta.ave.BR}, for a NH light neutrino spectrum. 

\begin{figure}
\begin{center}
\begin{tabular}{cc}
\epsfig{file=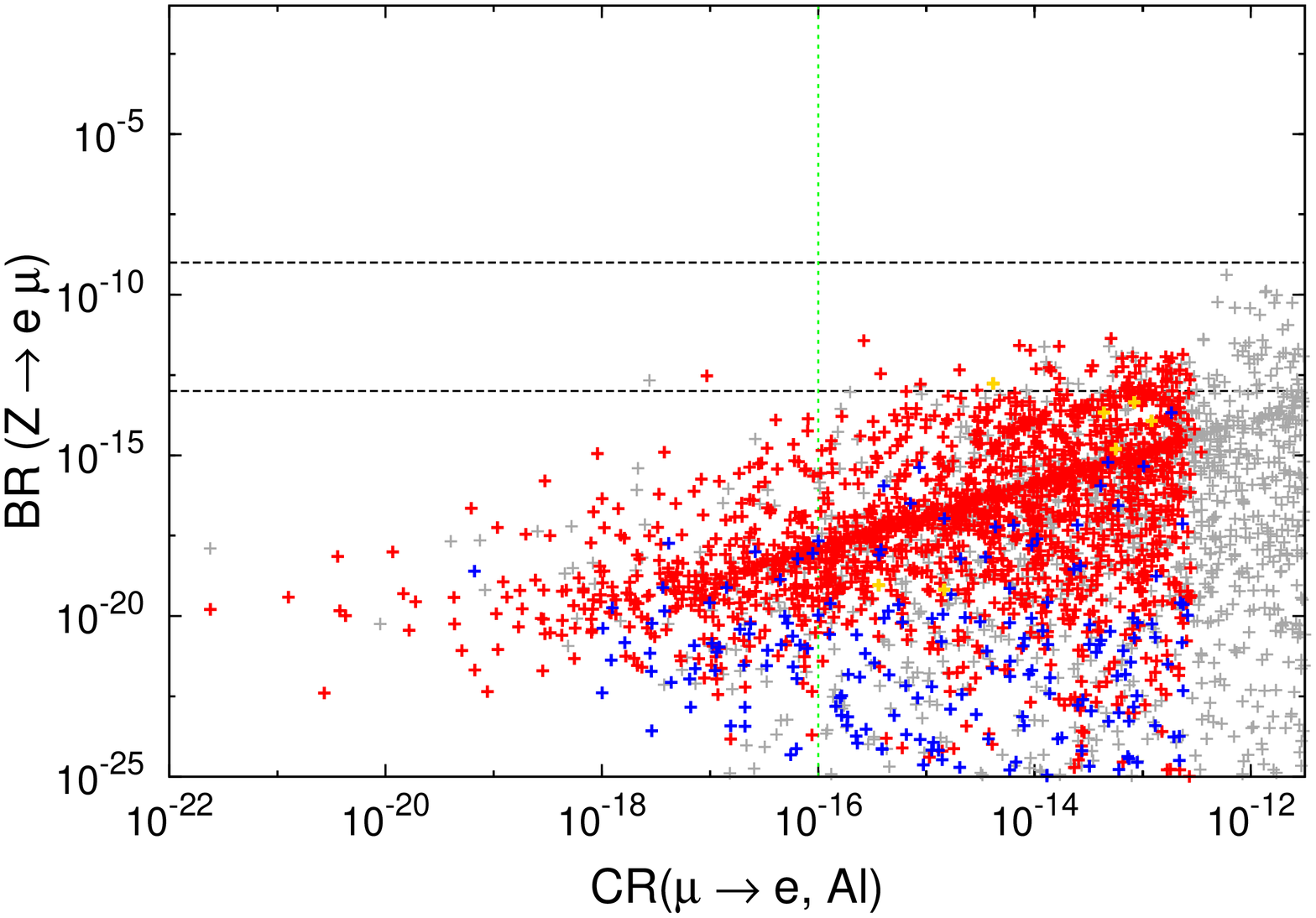, width=55mm, angle=270}
&
\epsfig{file=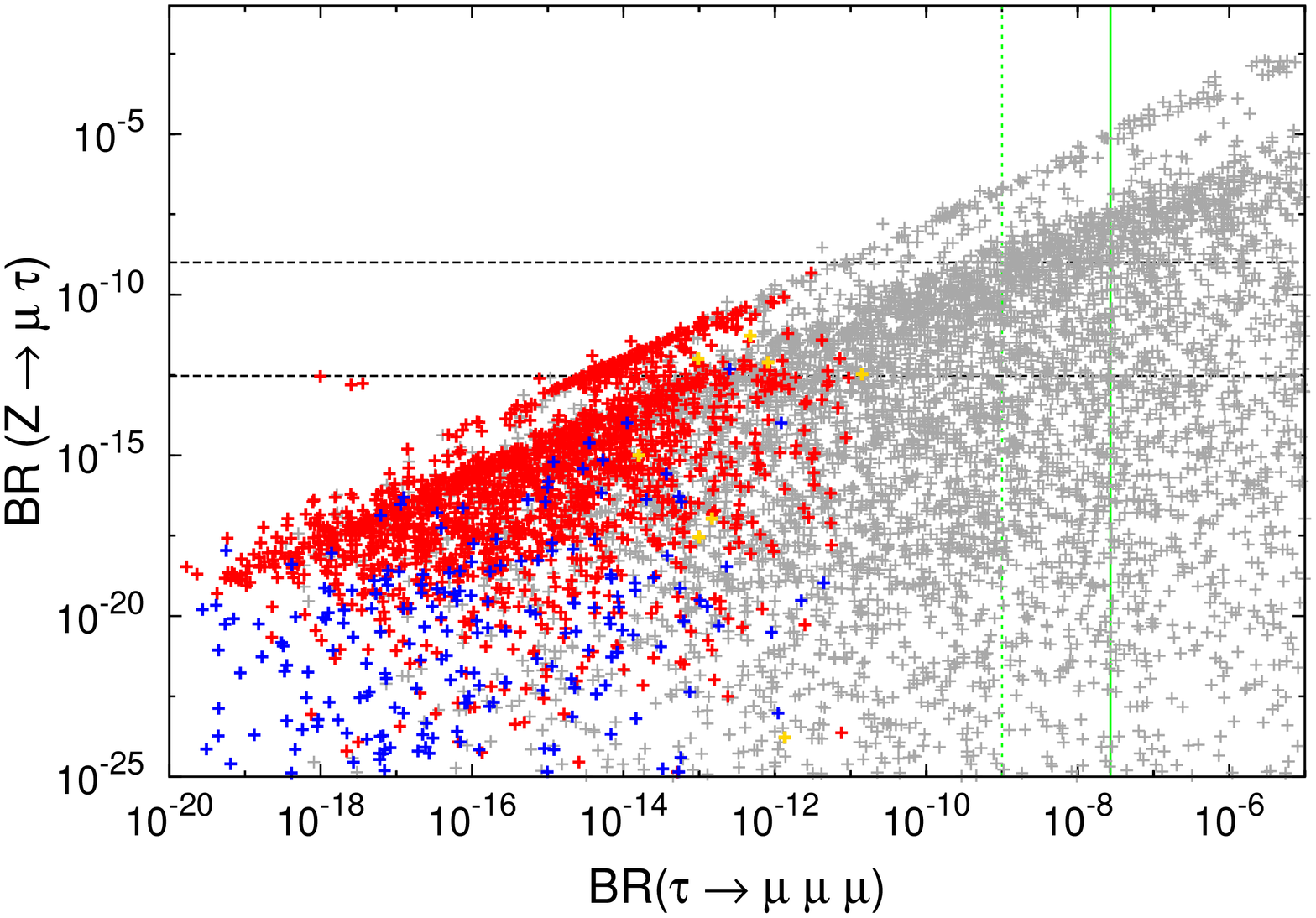, width=55mm, angle=270}
\end{tabular}
\end{center}
\caption{ISS realisation: on the left, BR($Z \to e \mu$) versus
  CR($\mu -e$, Al) and on the right BR($Z \to  \mu \tau$) versus
  BR($\tau 3 \mu$), for a NH light neutrino spectrum. 
  Line and colour code as in Fig.~\ref{fig:3+1:BRmt.m4}. 
  When present, the additional green vertical lines denote the current
  bounds (solid) and future sensitivity (dashed), and dark-yellow
  points denote an associated $|m_{ee}|$ within experimental reach.
}\label{fig:ISS:BRem.muegamma}
\end{figure}

The complementarity of low-energy LFV observables and 
LFV $Z$ decays at a high-luminosity $Z$ factory for
this ISS realisation is displayed in
Fig.~\ref{fig:ISS:BRem.muegamma}, where we further highlight points
that can potentially account for a $0\nu 2 \beta$ rate within
sensitivity of future experiments. The results are in agreement with
the findings for the ``3+1 model'': low-energy experiments - as COMET
looking for $\mu - e$ conversion in Al nuclei - are better probes of
cLFV in the $\mu - e$ sector of this (3,3) ISS realisation; on the other
hand, a future high-luminosity $Z$ factory has a stronger power 
to probe LFV in the $\mu-\tau$ sector via $Z$ decays.

\section{Overview}
In this work we have explored indirect searches for sterile neutrinos
at a future circular
collider running in the electron positron mode. In particular, we have
considered the impact of sterile neutrinos 
for (very) rare cLFV $Z$ decays, which can be
probed by the FCC-ee (TLEP) running close to the $Z$ mass
threshold, with an expected sensitivity to 
BR($Z \to \ell_1^\mp \ell_2^\pm$) as low as $10^{-13}$.

While these rare decays are forbidden in the SM (and have tiny BRs in its
ad-hoc extensions where neutrino masses and mixings are put by
hand), in models where the SM is extended via additional neutral sterile
fermions, which have non-negligible mixings with the active (light)
states, one can have significant contributions to cLFV $Z$ decays.

We have considered here three scenarios with sterile neutrinos: a
minimal ``3+1 toy model'', and two frameworks for neutrino mass
generation, the $\nu$MSM and the ISS. In our analysis we have 
conducted a thorough (numerical) exploration of the parameter space
of the different models: we take into account recent data on 
neutrino oscillations, as well as numerous experimental 
and observational constraints on the sterile states. 
As hinted by early analytical studies, and as a consequence of the
common LFV $Z \ell_1^\mp \ell_2^\pm$ vertex, low-energy cLFV observables
receiving contributions from $Z$-mediated penguins impose strong
constraints on the sterile neutrino induced BR($Z \to
\ell_1^\mp \ell_2^\pm$).  

The very minimal sterile extension of the SM - the ``3+1 model'' -
clearly illustrates the potential of the FCC-ee to probe the sterile
neutrino contributions to LFV $Z$ decays: both BR($Z \to \mu \tau$) and 
BR($Z \to e \tau$) are well within reach, especially for sterile masses
$\gtrsim 100$~GeV, and for sterile mixing angles $\theta_{i4} > 10^{-6}$.
Our analysis further emphasised the underlying synergy between 
a high-luminosity $Z$ factory and other dedicated (low-energy)
facilities: regions in ``3+1 model''
parameter space can be probed via cLFV $Z$ decays at FCC-ee,
through cLFV low-energy decays ($\tau \to 3 \mu$) and
neutrinoless double beta decays within reach of future dedicated
facilities (the latter especially in the case of an IH light neutrino 
spectrum); moreover, a high-luminosity $Z$ factory could probe LFV
in the $\mu-\tau$ sector, clearly going beyond the reach of low-energy
facilities.
Similar prospects were found for a (3,3) Inverse Seesaw realisation.
In contrast, the $\nu$MSM parameter 
space favoured by a successful generation of
the observed BAU turns out to be associated to very small values of 
BR($Z\to \ell_1^\mp \ell_2^\pm$), beyond the reach of the
FCC-ee. Nevertheless, direct searches for $\nu$MSM sterile states can
be carried at FCC-ee (for instance displaced vertices associated to long-lived
RH neutrinos~\cite{Blondel:2014bra}). 
We have also considered the violation of lepton flavour universality
in $Z$ decays, as encoded by the quatity
$\Delta R_Z^\text{lep}$. Still, in all the models here
considered, the estimated contributions of the sterile fermions to
this observable lie beyond experimental reach.

Our analysis reveals that sterile neutrinos can indeed give rise to 
contributions to BR($Z\to \ell_1^\mp \ell_2^\pm$)
within reach of the FCC-ee; these studies, in parallel with other
direct searches, have the potential to 
integrate the physics case of FCC-ee (TLEP).
Nevertheless, the results summarised here
consisted only of a first theoretical study:
a full discussion and estimation of the different
backgrounds, accompanied by simulations of the events and the
detector(s) will be required to ascertain whether or not one can
indeed have LFV signals above the background. This will be done in a
subsequent work.

\section*{Acknowledgements}
We are very grateful to  Y. Kuno for his helpful comments and
suggestions. We also thank S. Davidson for enlightening discussions.
The authors acknowledge support from the European Union FP7 ITN
INVISIBLES (Marie Curie Actions, PITN-\-GA-\-2011-\-289442). This work
was done in the framework of a ``PEPS PTI 2014'' project.

\appendix
\section{Loop integrals}
\label{loops}

The two- and three-point one-loop  dimensionless functions are defined as:
\begin{eqnarray}
\label{b1-def}
{\b}_1(x_i) &\equiv& B_1(0;m^2_i,M^2_W),
\\
\label{cbar-def}
\bar{\c}_{..}(x_i) 
&\equiv& M^2_W\ C_{..}(0,Q^2,0;m^2_i,M^2_W,M^2_W),
\\
\label{c-def}
\c_{..}(x_i,x_j) 
&\equiv& M^2_W\ C_{..}(0,Q^2,0;M^2_W,m^2_i,m^2_j),
\end{eqnarray}
from the usual loop integrals \cite{'tHooft:1972fi,Passarino:1979jh}
with the tensor decomposition (Minkowski metric):
\bea
\label{B-def}
B^\mu(p^2;m^2_0,m^2_1)&=&p^\mu B_1, 
\\
\label{cmu-def}
C^\mu(p^2_1,Q^2,p^2_2;m^2_0,m^2_1,m^2_2)&=&p^\mu_1 C_{11} + p^\mu_2 C_{12},
\\
\label{cmunu-def}
C^{\mu\nu}(p^2_1,Q^2,p^2_2;m^2_0,m^2_1,m^2_2)
&=&p^\mu_1 p^\nu_1 C_{21} + p^\mu_2 p^\nu_2 C_{22} 
+ (p^\mu_1 p^\nu_2 + p^\mu_2 p^\nu_1) C_{23}
+ g^{\mu\nu} M^2_W C_{24}. \nn\\
\eea

\end{document}